\documentclass{aa}
\usepackage{psfig}

\newcommand{\etal}{et al.}
\newcommand{\I}{\'{\i}}
\newcommand{\Ha}{\ifmmode {\rm H}\alpha \else H$\alpha$\fi}
\newcommand{\Hb}{\ifmmode {\rm H}\beta \else H$\beta$\fi}
\newcommand{\Hg}{\ifmmode {\rm H}\gamma \else H$\gamma$\fi}
\newcommand{\ha}{\ifmmode {\rm H}\alpha \else H$\alpha$\fi}
\newcommand{\hb}{\ifmmode {\rm H}\beta \else H$\beta$\fi}
\newcommand{\hg}{\ifmmode {\rm H}\gamma \else H$\gamma$\fi}
\newcommand{\whb}{\ifmmode EW({\rm H}\beta) \else $EW({\rm H}\beta)$\fi}
\newcommand{\wha}{\ifmmode EW({\rm H}\alpha) \else $EW({\rm H}\alpha)$\fi}
\newcommand{\hii}{H~{\sc ii}}

\newcommand{\nii}{N~{\sc ii}}
\newcommand{\oi}{O~{\sc i}}
\newcommand{\oii}{O~{\sc ii}}
\newcommand{\oiii}{O~{\sc iii}}
\newcommand{\sii}{S~{\sc ii}}
\newcommand{\siii}{S~{\sc iii}}

\newcommand{\heia}{He~{\sc i} $\lambda$5876} \newcommand{\oiiia}{[O~{\sc
iii}] $\lambda$4363}

\newcommand{\mstar}{$M_\star$}
\newcommand{\Ms}{\ifmmode {\rm M_\odot} \else M$_\odot$\fi}
\newcommand{\Zs}{\ifmmode {\rm Z_\odot} \else Z$_\odot$\fi}
\newcommand{\mup}{\ifmmode M_{\rm up} \else $M_{\rm up}$\fi}

\def\aap{A\&A}
\def\aas{A\&AS}
\def\aaps{A\&AS}
\def\aj{AJ}
\def\apj{ApJ}

\def\apjs{ApJS}
\def\apss{Ap\&SS}
\def\mnras{MNRAS}
\def\pasp{PASP}

\begin{document}

\thesaurus{08(11.01.1;11.05.2;11.09.4;11.19.3;11.19.5;09.08.01; 
  08.23.2)}

\title{The evolution of emission lines in \hii\ galaxies}
\author{ Gra\.{z}yna Stasi\'nska\inst{1}
\and Daniel Schaerer\inst{2}
\and Claus Leitherer\inst{3}}

\offprints{grazyna.stasinska@obspm.fr}
\institute{DAEC, Observatoire de Paris-Meudon, 92195 Meudon Cedex, France 
(grazyna.stasinska@obspm.fr)
\and
Laboratoire d'Astrophysique,
Observatoire Midi-Pyrenees, 14, Av. E. Belin, F-31400 Toulouse, France
(schaerer@ast.obs-mip.fr) \and
Space Telescope Science Institute\thanks{Operated by AURA for NASA under
contract NAS 5-26555}, 3700 San Martin Drive, Baltimore, MD 21218
(leitherer@stsci.edu)}
\date{Received 27 october 2000 / Accepted 13 february 2001}
\maketitle


\begin{abstract}

We constructed diagnostic diagrams using emission line ratios and
equivalent widths observed in several independent samples of \hii\
galaxies.
Significant trends are seen, both in the line ratio diagrams,
and in diagrams relating line ratios to the equivalent width of
\hb. The diagrams are compared to predictions from photoionization
models for evolving starbursts. This
study extends the work of Stasi\'nska \& Leitherer (1996) by including
objects with no direct determination of the metallicities, and
by using updated synthesis models with more recent stellar tracks and
atmospheres.

We find that \hii\ galaxies from objective-prism surveys are not satisfactorily
reproduced by simple models of instantaneous starbursts surrounded by constant density,
ionization bounded \hii\ regions.  The observed relations between
emission line ratios and \hb\ equivalent width can be understood if
older stellar populations generally contribute to the observed optical
continuum in \hii\ galaxies.  In addition, different dust obscuration
for stars and gas and leakage of Lyman continuum photons from the
observed \hii\ regions can be important.  As a result, \hii\ galaxies
selected from objective-prism surveys are not likely to contain
significant numbers of objects in which the most recent starburst is
older than about 5~Myr.  This explains the success of the strong line
method to derive oxygen abundances, at least in metal poor \hii\ 
galaxies.

The observed increase of [\oi]/\hb\ with decreasing \hb\ equivalent
width can result from the dynamical effects of winds and supernovae.
This interpretation provides at the same time a natural explanation of
the small range of ionization parameters in giant \hii\ regions.  The
classical diagnostic diagram [\oiii]/\hb\ vs [\oii]/\hb\ cannot be
fully understood in terms of pure photoionization models.  The largest
observed [\oii]/\hb\ ratios require additional heating.

The [\nii]/[\oii] ratio is shown to increase as the \hb\ equivalent
width decreases. A possible explanation is an N/O increase due to
gradual enrichment by winds from Wolf-Rayet stars on a time scale of
$\sim$ 5 Myr.  Alternatively, the relation between N/O and O/H could
be steeper than N/O $\propto$ O/H$^{0.5}$, with a previous stellar
generation more important at higher metallicities.

\end{abstract}

\section{Introduction}

Isolated extragalactic \hii\ regions, which we will simply refer to as
\hii\ galaxies, are powered by clusters of hot, massive stars that ionize
their environment. Recent analyses of such objects, comparing the
observed nebular emission lines, colors, and stellar features with
evolutionary models (Mas-Hesse \& Kunth 1991, 1999, Schaerer et al.\ 1999,
Stasi\'nska \& Schaerer 1999) found that most have
experienced a recent, quasi instantaneous burst of star formation.
In addition, there is growing evidence for an older
stellar population (Telles \& Terlevich 1997, Papaderos et al.\ 1998,
Raimann et al.\ 2000a).  Some clues regarding the evolution of giant \hii\ 
regions, their properties and stellar content may be obtained by
considering \hii\ galaxies in different evolutionary stages with
respect to the most recent star formation event.

The present paper follows this approach: we compare the emission line
properties of a large sample of \hii\ galaxies with those of
photoionization models in which the ionizing continuum is provided by
synthetic models of evolving starbursts. In a previous paper (Stasi\'nska
\& Leitherer 1996, hereinafter SL96), we considered a sample of metal poor
galaxies in which the oxygen abundance could be derived directly from
observations using the electron temperature sensitive \oiiia\ line. Thus
the derived abundances were unambiguous and model independent.
Unfortunately, we were automatically restricted to young starbursts by 
this requirement. As the most massive stars disappear, the
radiation field gradually softens inducing a decrease of the [\oiii]
emission, and the weak \oiiia\ becomes undetectable after about 5~Myr.
Alternatively, the oxygen abundances can be estimated via strong line methods
(Pagel et al.\ 1979, McGaugh 1991, 1994). The relevance and accuracy of these
methods is however still under debate. We address this question in Sect. 4.2.

Lifting the restriction to the youngest starbursts immediately leads
to inclusion of objects with a priori unknown metallicities.  However,
the distribution in metallicities should be the same for old and young
objects, permitting an interpretation of diagnostic diagrams based on
emission lines.  This advantage of a large sample of bona fide young
and old \hii\ galaxies is explored in the present work. With respect
to SL96, we also use updated evolutionary synthesis models with more
recent stellar tracks and stellar atmospheres as input for our
photoionization models.

The study in the present paper is complementary to work on giant \hii\ 
regions in spiral galaxies, such as published by Garc\I a-Vargas \&
D\I az (1994), Garc\I a-Vargas et al.\ (1995), and more recently
Bresolin et al.\ (1999) and Dopita et al.\ (2000).  First, our
observational samples are based on extragalactic \hii\ regions, for
which the equivalent width of \hb\ can be used as a direct first-order
indicator of the age of the latest starburst (Dottori 1981).  Second,
our samples are biased towards dwarf galaxies with correspondingly low
average metallicities.

In Section~2 we present the observational samples, discuss possible
selection effects, and display a series of observational diagrams. In
particular, we show that emission line trends are seen not only in
classical emission line ratio diagrams, but also in diagrams relating
emission line ratios with the \hb\ equivalent width.  In Section~3 we
describe the models, present our model grid, and show theoretical
diagrams for evolving starbursts. In Section~4 observations and models
are compared.  The main conclusions are summarized in Section~5.

\section{Sample selection}

\subsection{The database}

The prerequisites for our observational data base are the following.
We require a large sample of objects with uniform observation
parameters.  The sample must be large since there are several factors
determining the line spectra, the most important being metallicity and
starburst age. We wish to use information from the \heia\ line, which
is directly related to the mean effective temperature of the ionizing
stars (cf. Stasi\'nska 1996).  Deep spectroscopy is required for this
generally weak line.  In addition, we want to make use of a direct
estimator of the starburst age, such as provided by the equivalent
width of \hb\ (hereafter \whb).  While this estimator has its
well-known drawbacks (effects of dust, incomplete absorption of the
stellar ionizing photons by the observed nebula), it is relatively
independent of the assumed nebular properties, as compared with
emission line ratios such as [\sii] $\lambda\lambda$6717+30/\hb\ or
[\sii] $\lambda\lambda$6717+30/[\siii] $\lambda$9069 proposed by
Garc\I a-Vargas et al.\ (1995). However, \whb\ is difficult to use for
giant \hii\ regions with a strong underlying old stellar component,
such as in nuclear starbursts or giant \hii\ regions in evolved
galaxies. Our sample is thus restricted to giant \hii\ regions in
irregular or blue compact galaxies, where the old stellar component is
expected to be weaker.

The largest, homogeneous sample meeting these requirements is the
spectrophotometric catalogue of \hii\ galaxies by Terlevich et al.\ 
(1991). It contains spectra of 425 emission line galaxies, among which
about 80\% are classified as genuine \hii\ galaxies by the authors (as
opposed to active galactic nuclei and nuclear starbursts). Further
imaging of these objects identified some of these \hii\ galaxies as
giant \hii\ regions in spiral galaxies (Telles, private
communication). Those objects were removed to produce the final
sample, hereafter referred to as the Terlevich sample, which consists
of 305 objects. Most of the observations for the Terlevich catalogue
were made with 2-m class telescopes, and the typical signal-to-noise
in the continuum is 5. Note, however, that second-order contamination
could affect some of the line ratios for $\lambda >$~6000 \AA\ 
(Terlevich et al.\ 1991).

As a control sample, we consider a much more restricted collection of
isolated extragalactic \hii\ regions with very high quality
observational data. These are the blue compact galaxies observed by
Izotov and coworkers (Izotov et al.\ 1994, 1997, Thuan et al.\ 1995,
Izotov \& Thuan 1998, Guseva et al.\ 2000) on 2-m and 4-m class
telescopes. Their signal-to-noise in the continuum is typically
20~--~40.  The latter sample, referred to as the Izotov sample,
comprises 69 objects.

Recently, another survey of emission line galaxies meeting our
requirements has been published. This is the catalogue of Popescu \& Hopp
(2000) which contains 90 objects, out of which 70 are classified as \hii\
galaxies. The quality of the data is not as good as for the Izotov sample,
but the sample is more homogeneous, as explained below.

\subsection{Selection effects}

Objective-prism surveys of emission line objects tend to underestimate
the true proportion of objects with weak emission line equivalent
widths. On the other hand, genuine, isolated giant extragalactic \hii\ 
regions will always be discovered in such surveys if they are bright
enough. The Terlevich catalogue comprises 75\% (25\%) of the
non-quasar emission line objects identified in the Michigan (Tololo)
objective-prism survey, and additional biases are not expected to be
important. Therefore, one expects a roughly uniform distribution of
ages of the most recent starbursts, and the same distribution of
metallicities at each age bin above a certain threshold in \hb\ 
equivalent width in the Terlevich sample.

In the case of the Izotov sample, the objects are mainly blue compact
galaxies from the First and Second Byurakan objective-prism surveys.
The sample is obtained by merging two sub-samples. One is composed of
objects which had been selected for their low metallicities ($Z <
\Zs/10$) as estimated from low signal-to-noise 6-m telescope spectra.
The other subsample (the 39 objects studied by Guseva et al.\ 2000) is
composed of objects selected on the basis of broad Wolf-Rayet features
seen in the 6-m telescope spectra. The metallicities in the latter
subsample range between \Zs/40 and \Zs\footnote{For 9 objects the
  electron temperature was too low for temperature sensitive ratios to
  be measured, and the metallicity was estimated with the strong line
  method (Pagel et al.\ 1979).}. The \hb\ equivalent widths in the
Guseva subsample are, on average, lower than those in the other Izotov
et al.\ subsample, and contamination from old stellar populations is
likely present at least for some objects known as nuclear starbursts.
Compared to the Terlevich sample, the Izotov sample is affected by
strong biases which must be kept in mind when interpreting the
diagnostic diagrams shown below.

The Popescu \& Hopp sample is a complete subsample of emission line
galaxies found on the objective-prism plates for the Hamburg Quasar
Survey. The single selection criterion was the location of the galaxies with
respect to voids. Therefore, as for the Terlevich sample, we expect a
roughly uniform distribution of ages, and the same metallicity
distribution at each age bin. The sample is however much smaller than the
Terlevich sample; therefore  statistical fluctuations may not be
negligible.

\subsection{Reddening corrections}

Izotov and coworkers used an iterative procedure to deredden their
sample where the reddening constant and stellar absorption at \hb\ are
determined simultaneously. Such a procedure was feasible due to the
high signal-to-noise of the data. They assumed the Whitford (1958)
extinction law and took for the intrinsic \ha/\hb\ ratio the
recombination value corresponding to the electron temperature derived
from [\oiii] 4363/5007, when available. In most cases, the
contribution of the stellar absorption in \Hb\ turns out to be only a
few percent, except in some objects from the Guseva sample which
includes a few nuclear starbursts (see Guseva et al.\ 2000, Schaerer
et al.\ 2000).

The line intensities in the Popescu \& Hopp (2000) sample are
published corrected for reddening. For the galaxies with a strong
underlying continuum relative to the \Hb\ emission (\whb\ $<$ 20 \AA),
they assumed a constant underlying absorption at \Hb\ of 2~\AA. The
reddening was computed from \ha/\hb\ assuming an intrinsic ratio of
2.87 and using the Howarth (1983) extinction law.

In the case of the Terlevich sample, the data are published without
reddening correction. We have dereddened them adopting the Seaton
(1979) reddening law and an intrinsic \Ha/\Hb\ ratio of 2.87. We
assumed a constant underlying stellar absorption of 3~\AA\ for all the
spectra in the \Ha\ and \hb\ lines, which is predicted by models for
synthetic absorption line spectra in starburst galaxies (Olofsson
1995a, Gonz\'alez Delgado et al.\ 2000). Unfortunately, spectra in the
\Ha\ region have second order contamination in many objects (cf.\ 
Terlevich et al.\ 1991), affecting the reddening correction.
Alternatively, one could use \Hg/\hb\ instead of \Ha/\hb\ to determine
the reddening correction, but this procedure is plagued by the fact
that \Hg\ is a much weaker line than \Ha, introducing substantial
errors due to the reduced signal-to-noise, and by the lack of
published \Hg\ equivalent widths. In any event, for objects without
published \Ha\ data, we used \Hg/\hb\ to derive the reddening,
assuming that the stellar (negative) contribution to the observed \Hg\ 
line is three times as large as for the \hb\ line.

The reddening correction introduces some uncertainties into each
sample.  Even in the case of the Izotov sample, where dereddening was
performed in a completely self-consistent way, the reddening
correction may not be perfect.  First, departures from the standard
reddening law exist (e.g.\ Mathis \& Cardelli 1988 or Stasi\'nska et
al.\ 1992). Second, the intrinsic \Ha/\hb\ ratio in a nebula is not
necessarily the recombination value. In objects with high electron
temperatures, collisional excitation increases \Ha/\hb\ with respect
to the recombination value, in a proportion that depends on the
quantity of neutral hydrogen in the emitting regions (Davidson \&
Kinman 1985, Stasi\'nska \& Schaerer 1999).  Adopting the same
underlying absorption of 3 or 2~\AA\ at \Hb\ for all the objects of
the Terlevich and Popescu \& Hopp samples may not be justified if
stellar populations from previous star-forming events contribute
significantly to the continuum (see below).  Among the line ratios
discussed in this paper, the one most affected by reddening is
[\oii]/\Hb. It is prudent to assume that the [\oii]/\hb\ ratios may
well be uncertain by up to about 30\%, and even more for some objects
in the Terlevich sample.

\subsection{Observed emission line trends}

The nine panels of Figs.\ 1, 2, and 3 are various observational
diagrams for the Terlevich, Izotov, and Popescu \& Hopp samples,
respectively.  Panels a through f in these figures show the behavior
of various emission line ratios as a function of \whb.  In spite of
some (expected) dispersion, striking line trends are seen. These
trends are very similar in the three samples, and are shown for the
first time.

Some differences among the samples are likely due to the quality of
the data. The smaller dispersion in \heia/\hb\ in the Izotov sample,
for example, is probably due to the better signal-to-noise. We also
note that the Terlevich and Popescu \& Hopp samples contain some
objects with rather large [\oii]/\Hb: $\sim$ 20 \% of the Popescu \&
Hopp sample and 3\% of the Terlevich sample have [\oii]/\hb\ $\ge$5,
while no object of the Izotov sample has such large values for this
ratio.  However, if we retain only the 50 objects from the Terlevich
sample with measurements in the four lines [\oii] $\lambda$3727,
[\oiii] $\lambda$5007, \heia\ and [\nii] $\lambda$6584, the diagrams
become very similar to those corresponding to the Izotov sample.
Therefore, we will ignore the objects with [\oii]/\hb\ $>$ 5 since
such high ratios may be attributed to poor signal-to-noise and/or to
inadequate reddening corrections.

We generally note a gradual increase in the dispersion and a decrease
of the average value of [\oiii]/\hb\ as \whb\ decreases, while the
\heia/\hb\ ratio remains remarkably constant. The [\oi]/\hb\ ratio, on
the other hand, steadily increases with decreasing \whb, extending the
trend already noted by SL96 to lower \whb. The [\oii]/\hb\ increases
and tends to level off at \whb\ about 30 \AA. The ([\oii]+
[\oiii])/\hb\ ratio is rather constant at the largest \whb, and
becomes more dispersed with a tendency to decrease as \whb\ decreases.
The [\nii]/[\oii] ratio shows a clear tendency to increase as \whb\ 
decreases, although the dispersion is larger than in the remaining
diagrams. Overall, all the trends of emission line ratios with \hb\ 
equivalent width are significant and require an explanation.

Finally, panels g, h, and i in Figs.\ 1 -- 3 show the standard
emission line ratio diagrams ([\oiii]/\hb\ vs [\oii]/\Hb, [\oiii]/\hb\ 
vs [\sii]/\hb\ and [\nii]/[\oii] vs ([\oii]+[\oiii])/\Hb), similar to
those that have been widely used in the literature to distinguish
\hii\ regions from active galaxies (Baldwin et al.\ 1981, Veilleux \&
Osterbrock 1987, van Zee et al.\ 1998, Martin \& Friedli 1999) and for
studies of abundance ratios in \hii\ regions (McGaugh 1994, Ryder
1995, Kennicutt \& Garnett 1996, van Zee et al.\ 1998, Bresolin et
al.\ 1999). For comparison, we have plotted in Fig.\ 4 the same three
emission line ratio diagrams for the sample of giant \hii\ regions in
{\it spiral} galaxies from McCall et al.\ (1985).

The \hii\ galaxies of our samples clearly appear like an extension of
the giant \hii\ region sequence of McCall et al.\ toward the high
[\oiii]/\hb\ end and towards the low [\nii]/[\oii] end.  Some of the
\hii\ galaxies (mainly in the Terlevich sample) reach values of
[\nii]/[\oii] which are as high as those observed in the giant \hii\ 
regions of the central parts of spiral galaxies. The true scatter in
[\oii]/\hb\ is probably smaller than it appears in our figures: as
discussed in Section~2.3, [\oii]/\hb\ is strongly affected by
uncertainties in the dereddening procedure\footnote{ Plotting
  [\oiii]/\hb\ vs [\oii]/\hb\ from the Terlevich sample using line
  ratios not corrected for reddening results in a smaller scatter.}.
Note the pronounced bending of the [\oiii]/\hb\ vs [\oii]/\hb\ and
[\oiii]/\hb\ vs [\sii]/\hb\ sequences toward the left at high
[\oiii]/\hb\ in the three \hii\ galaxies samples. The relation between
[\nii]/[\oii] and ([\oii]+[\oiii])/\hb\ is extremely tight in all the
samples we consider. This was also seen in diagrams originally studied
by the authors mentioned above. Considering that line ratios in
nebulae are determined by several factors (abundance ratios, intensity
and spectral distribution of the ionizing radiation field, density
distribution of the nebular gas), the question arises as to why the observed
sequences are so narrow.

We will try to understand the observational data in the
light of the photoionization models described in the next section.

\section{The models}

\subsection{Technique and assumptions}
\label{s_models}

The evolutionary synthesis models of Schaerer \& Vacca (1998) are used
to predict the theoretical spectral energy distributions\footnote{The
  predictions from these synthesis models are available on the Web at
 http://webast.ast.obs-mip.fr/people/schaerer/.}.  They are
based on the non-rotating Geneva stellar evolution models, with the
high mass-loss tracks of Meynet et al.\ (1994).  The spectral energy
distributions for massive main-sequence stars are those given by the
{\em CoStar} models (Schaerer \& de Koter 1997) taking into account
the effects of stellar winds, non-LTE, and line blanketing.  The pure
He models of Schmutz et al.\ (1992) are used for Wolf-Rayet stars.
The spectral energy distributions from the plane-parallel LTE models
of Kurucz (1991) are used for the remaining stars which contribute to
the continuum.

To illustrate the differences between the ionizing spectra adopted here
with the recent {\em Starburst99} models (Leitherer et al.\ 1999) and
the earlier Leitherer \& Heckman (1995) models adopted in SL96, we
plot the ratio of the He$^0$ and He$^{+}$ ionizing photons ($Q_{\rm
  He}$, $Q_{\rm He^+}$ respectively) to $Q_{\rm H}$, the number of H
ionizing photons issued from various models in Figure \ref{fig_q1q0}.
Differences with the Leitherer \& Heckman (1995) models (left panel)
are mostly due to the updated stellar tracks, which in particular,
lead to a shorter Wolf-Rayet (WR) phase in the solar metallicity model
shown, and thus to a more rapid decline of the hardness of the
ionizing spectrum.  From this figure it is also apparent that at solar
(and higher) metallicities the changes brought about by recent stellar
tracks have a rather important impact on the ionizing fluxes, which
affects photoionization models based on the older Leitherer \& Heckman
(1995) spectra (e.g.\ SL96, Bresolin \etal\ 1999).
As both the Schaerer \& Vacca (1998) and {\em Starburst99} models use
the same stellar tracks, the differences (right panel) in the
predicted spectra are only due to different treatments of stellar
atmospheres.  Outside the phases with WR stars (i.e.\ at ages $\la$ 3
Myr or $\ga$ 6 Myr), the Schaerer \& Vacca spectrum in the He$^{0}$
ionizing continuum is somewhat harder due to the use of improved O
star model atmospheres (cf.\ also Schaerer \& Vacca 1998, Fig.\ 5).
Differences in the He$^+$ ionizing flux (non-negligible during part of
the WR phase; cf.\ Schaerer \& Vacca) are due to slightly different
assumptions on the link between the stellar evolution and atmosphere
models and other numerical differences.

As also apparent from Fig.\ \ref{fig_q1q0}, WR stars are predicted to
provide a non-negligible fraction of the ionizing flux, especially at
high energies during the WR-rich phase, whose duration increases with
metallicity (cf.\ Schaerer \& Vacca 1998). The predicted ionizing
fluxes are likely more uncertain during these phases, due to 1) the
neglect of metals in the WR atmosphere models of Schmutz \etal\ 
(1992), and 2) uncertainties in relating the stellar interior and
atmosphere models (cf.\ review by Schaerer 2000). However, while the
predicted spectra are possibly too hard at high metallicities (see
also Crowther \etal\ 1999, Bresolin \etal\ 1999), there are good
indications that at low metallicities -- of main concern here -- the
non blanketed WR atmospheres combined with the latest stellar tracks
provide a reasonable description of the ionizing spectra (e.g.\ de
Mello \etal\ 1998, Luridiana \etal\ 1999, Gonz\'alez Delgado \&
P\'erez 2000, Schaerer 2000 ).

We used the code PHOTO to build the photoionization models, with exactly
the same atomic data as in SL96.

\subsection{The model grid}

We built a wide model grid in order to cover the full parameter space
found in our observational samples\footnote{Tables giving infrared,
  optical and ultraviolet line intensities for all the models
  discussed in this paper are available on the Web at
  http://webast.ast.obs-mip.fr/people/schaerer/}.

Most models are computed for instantaneous bursts with a Salpeter
initial mass function and an upper stellar limit \mup\ = 120 \Ms. The
evolution of the burst is followed in time steps of 10$^{6}$ yr,
starting from an age of 10$^{4}$ yr. Such a time step may be too long
for a detailed description of the Wolf-Rayet stage in the evolution of
the burst at metallicities lower than \Zs/5 (see Schaerer \& Vacca
1998), but is a good compromise to cover the evolution until 10~Myr. A
second set of models was computed for an upper stellar limit of
30~\Ms\ instead of 120~\Ms. Another set of models was built with \mup\ 
= 120 \Ms\ and a Salpeter IMF but for a star formation extended over a
period of 10~Myr.

We considered five metallicities: 2, 1, 0.4, 0.2 and 0.02 times the solar
metallicity. The metallicities of the stars and the nebular gas are assumed
to be the same. As in SL96, the metallicity is defined as the oxygen
abundance (oxygen is the main gas coolant). O/H is taken equal to $8.51
\times 10^{-4}$ for the solar abundance set. The abundances of the other
elements relative to oxygen are given by the prescription of McGaugh
(1991). In particular, He/H = 0.0772 + 15 (O/H) and $\log$ N/O = 0.5
$\log$ (O/H) + 0.4. The ionized gas is assumed to be dust free.

The third main parameter that determines the emission line ratios and
equivalent widths of a model is the gas distribution. The
photoionization models in our grid are spherically symmetric with
constant density, and are ionization bounded. Most models presented in
this paper are spheres uniformly filled with gas at a density of $n$ =
10~cm$^{-3}$. We also constructed some models represented by hollow
spheres with $f=3$ and $n$ = 200 cm$^{-3}$, where $f$ is the inner
radius in units of Str\"omgren radius for a full sphere (see Stasi\'nska
\& Schaerer 1997). These two extreme cases are intended to highlight
the sensitivity of the results on the adopted density distribution. It
can easily be shown that such hollow sphere models have the same
ionization parameter (averaged over the nebular volume) as the filled
sphere models with $n$ = 10 cm$^{-3}$ and a total ionizing luminosity
1000 times lower.

Four values of the ionization parameter are considerered: those that
would be produced by a star cluster of \mstar= 1, 10$^{3}$, 10$^{6}$,
and 10$^{9}$ \Ms\footnote{A lower IMF mass cut-off of $M_{\rm low}$ =
  0.8 \Ms\ was adopted.}, providing a range in ionization parameters
of a factor 1000 at a given epoch. One should realize that our choice
of density distribution is quite arbitrary. So far, little is known
theoretically or observationally about the systematics of the
evolution of the nebular geometry with time in giant \hii\ regions.
Our model sequences are not necessarily meant to describe the true
evolution of a giant \hii\ region with time. For example, it may well
be that, due to the combined effects of expansion or stellar winds,
the ionization parameter of giant \hii\ regions decreases with time
more than by the sole effect of the decrease of the number of ionizing
photons. This was suggested by SL96 from the comparison of a sample of
young \hii\ galaxies with their grid of models. Furthermore, the time
variation of the ionization parameter depends on the adopted evolution
of the geometry. For example, the ionization parameter $U$ in our
models varies as $Q_{\rm H}^{1/3}$ ($Q_{\rm H}$ being the number of
stellar photons able to ionize hydrogen per unit time) while in the
models of Garc\I a-Vargas \& D\I az (1994) and Garc\I a-Vargas et al.\ 
(1995), who constructed shells with a fixed outer nebular radius, $U$
varies as $Q_{\rm H}$.

Figures 6,7 and 8 correspond to our full sphere model sequences for
instantaneous bursts with \mup\ = 120 \Ms\ and \mstar\ = 10$^{9}$,
10$^{6}$, and 10$^{3}$ \Ms, respectively. Figure~9 corresponds to a
\mstar\ = 10$^{6}$ \Ms\ burst uniformly extended over a period of 10
Myr. (In this case the time step is 2 Myr instead of 1 Myr.) Figure~10
is for a \mstar\ = 10$^{6}$ burst with \mup\ = 30 \Ms. Figure~11
corresponds to our hollow sphere model sequences with \mup\ = 120 \Ms\ 
and \mstar\ = 10$^{9}$ \Ms.  Each model is represented by a symbol:
circle for Z/\Zs\ = 2, square for Z/\Zs\ = 1, triangle for Z/\Zs\ =
0.4, diamond for Z/\Zs\ = 0.2 and plus sign for Z/\Zs\ = 0.05, and the
successive models in a sequence are linked by a thin line.

It is important to realize that, while the model predictions for
optical forbidden lines like [\oiii] $\lambda$5007, [\oii]
$\lambda$3727, [\nii] $\lambda$6584 are relatively robust and accurate
for metallicities below solar, the situation worsens as one goes to
higher metallicities. The reason is that, as metallicity increases,
cooling by heavy elements becomes more and more important, and is
gradually shifted from the optical forbidden lines to the fine
structure, infrared lines, which have lower excitation potential.  The
emissivities of the infrared lines are little dependent on the
electron temperature. As a consequence, the computed value of the
electron temperature, obtained by equating the energy gains and losses
of the electrons, strongly depends on the computed energy loss.
Therefore, any small variation in the energy loss will induce an
appreciable change in the emission of the optical forbidden lines. For
example, since the infrared fine structure lines are easily
collisionally de-excited, any gas clump of density above 500 cm$^{-3}$
will induce collisional de-excitation of the infrared cooling lines,
boosting the [\oiii] $\lambda$5007 line. Another problem is that in
many cases photoionized gas at temperatures below $\sim$ 5000~K is
thermally unstable.  Thus the temperatures computed in photoionization
models do not reflect the whole range of temperatures that can be
found in real objects.  For these reasons the predicted intensities of
the models with Z/\Zs\ $>$ 1 (corresponding to O/H = $8.51 \times
10^{-4}$ as stated above) are less accurate than those of lower
metallicity models. Another notorious problem of high metallicity
models is that the intensities of optical forbidden lines, especially
from the low ionization species, are very dependent on the amount of
metals trapped in dust grains (Henry 1993, Shields \& Kennicutt 1995).
Our reference grids of models assume no depletion, although we have
constructed additional models with only one tenth of the metals in the
gas phase, see Sect.~4.3)

\section{Observations versus models}
We first analyze diagnostic diagrams involving the \hb\ equivalent
width. The classical, pure emission line ratio diagnostic diagrams
will be discussed subsequently.

\subsection{Diagnostic diagrams with equivalent widths}

The observed samples of \hii\ galaxies, and giant \hii\ regions in
general, are characterized by a very narrow range in ionization
parameters. This was already noted before from a comparison of
observed diagrams with model grids (McCall et al.\ 1985, Dopita \&
Evans 1986, Garc\I a-Vargas \& D\I az 1994, Garc\I a-Vargas et al.\ 
1995, Stasi\'nska \& Leitherer 1996, Bresolin et al.\ 1999, Dopita et
al.\ 2000). We show in Fig.\ 12 the values of [\oiii]/[\oii] as a
function of \whb\ for models with \mstar\ = 1, 10$^3$, 10$^6$ and
10$^9$~\Ms\ and metallicity 0.2~\Zs\ (corresponding to a range of a
factor 10 in $U$ between each curve) superimposed on the Terlevich
data.  For \whb\ $>$ 100 \AA, the observed dispersion in
[\oiii]/[\oii] suggests a range in ionization parameters of about a
factor 10. Most of the data are bracketed by the series with \mstar =
10$^3$ and 10$^6$ \Ms\ (corresponding roughly $U$ $\sim$
$1-2\times10^{-3}$ and to $U$ $\sim$ $1-2\times10^{-2}$, for the first
5 Myr).  At lower equivalent widths, the observations fall outside the
predicted values, apparently indicating a much higher $U$. As we will
show below, this would be a premature conclusion.

If \hii\ galaxies are powered by coeval star clusters, one expects the
[\oiii] $\lambda$5007/\hb\ ratio to decrease as time goes by and \whb\ 
decreases.  Panels a in Figs.\ 6-11 show the predictions for [\oiii]
$\lambda$5007/\hb\ as a function of \whb.  All the sequences show a
steep decline of [\oiii] $\lambda$5007/\hb\ at an \whb\ around 20 \AA,
irrespective of metallicity. A similar effect was predicted by the
models of SL96. This is due to the average effective temperature of
the exciting stars (as roughly expressed by $Q_{\rm He^0}/Q_{\rm H}$)
dropping much more rapidly than the total number of ionizing photons,
once the most massive stars have disappeared. The rise in the curves
in Figs.\ 6--11 prior to the final steep decline for the $Z=2$~\Zs\ 
and \Zs\ models is due to the appearance of Wolf-Rayet stars.  The
sequences of models with \mup\ = 30 \Ms\ show a strong accumulation of
points at early ages since it takes stars with masses below 30 \Ms\ 
more than 4~Myr to evolve off the main-sequence. The drop in [\oiii]
$\lambda$5007/\hb\ at \whb\ smaller than 20~\AA\ is of course the same
as for the \mup\ = 120 \Ms\ sequences. The sample of \hii\ galaxies in
the SL96 study had \whb\ too large to show this drop, as a consequence
of the requirement on \oiiia, as stated in the introduction. Since no
such restriction was applied here, the values of \whb\ go down to a
few Angstroms. The data in each sample have a tendency to a decrease
of [\oiii] $\lambda$5007/\hb\ with decreasing \whb. This decrease,
however, is rather mild and the scatter in [\oiii] $\lambda$5007/\hb\ 
increases for lower \whb.  One might think that extending the star
formation over a longer period would help
to reconcile the predicted trends with the observations.
However, Fig.\ 9 suggests that even considering a constant star formation
during 10~Myr is not sufficient to produce the large [\oiii]/\hb\ observed
at \whb\ of 20 \AA\ or smaller.

A similar effect is seen in the \heia/\hb\ versus \whb\ diagrams
(panels b). In contrast to [\oiii]/\hb\ which, for a given stellar
radiation field depends on $U$ and $Z$, \heia/\hb\ is independent of
the ionization parameter and depends only weakly on metallicity (in
the sense that at larger metallicities a lower electron temperature
increases \heia/\hb). Instantaneous starburst models predict a sharp
drop in \heia/\hb\ at \whb\ around 20 \AA. This is not seen in any of
the observed samples. The observed \heia/\hb\ ratio in the Terlevich
sample remains remarkably constant around a value of 0.1.  The slight
rise and the increasing dispersion at low \whb\ could be due to
observational errors, \heia\ being a weak line. Unfortunately, this
line is measured in only 98 out of 305 objects, and one might argue
that the remaining objects have a much lower \heia/\hb\ ratio.  
{\em All} the objects in the Izotov sample have the \heia\ intensity
measured with very good accuracy, and all except one are of the order
of 0.1.  Although the statistics are very poor, since only 10 objects
in the Izotov sample have \whb\ of the order of 20 \AA\ or smaller,
there seems to be no drop of \heia/\hb\ at this \whb. The Popescu \&
Hopp sample shows the same behavior as the Terlevich sample.

The presence of an older underlying stellar component in \hii\ 
galaxies, as supported by many studies based on infrared colors and
stellar features, changes the interpretation of the diagrams.  It
would decrease the \hb\ equivalent width of the young, ionizing
population (``dilution'' effect), and also lead to an underestimate of
the nebular \hb\ luminosity due to underlying stellar absorption.
Qualitatively, these effects work in the direction required to
reconciliate models and observations.  Such an explanation is
suggested by a recent study of Raimann et al.\ (2000a). These authors
grouped 185 spectra from the Terlevich et al.\ (1991) paper into 19
templates in order to increase the signal-to-noise and to allow a
population synthesis analysis from the absorption features. They found
that \hii\ galaxies are age-composite stellar systems, and that they
can contain a significant population of stars with ages larger than~10
Myr.  In Fig.~13 we plotted the same quantities as in
Figs.~1--3\footnote {except \heia/\hb\ which is not available} for the
spectral groups defined by Raiman et al.\ (2000a) (excluding the 3
groups corresponding to Seyferts which are not discussed in the
present paper). Open circles represent the averaged observed spectra.
Each open circle is linked by a straight line with a filled circle
representing the data after correction for an underlying stellar
population (line and continuum) and for internal reddening.  The same
trends with \whb\ are seen for the filled circles as for the data
shown in Figs. 1--3, but with larger slopes. The drop in [\oiii]/\hb\ 
starts at much higher values of \whb. This is much more compatible
with our models than the uncorrected data. The steep decline starts
immediately, which seems to be in better agreement with our models
with \mup = 30 \Ms (Fig.~10) than with 120~\Ms. However, the
distribution of the open circles in the [\oiii]/\hb\ versus \whb\ 
diagram of Fig.~13 is quite different from the distribution of the
individual galaxies from the Terlevich sample (Fig.~1); therefore this
averaged sample is unsuitable for describing the general properties of
\hii\ galaxies. The important point is that the work of Raimann et
al.\ (2000a, b) demonstrates that signatures of older underlying
stellar populations are present in the spectra of \hii\ galaxies and
can explain the observed trend in the uncorrected [\oiii]/\hb\ versus
\whb\ diagrams of Figs.~1--~3. The multiwavelength analysis of 17
\hii\ galaxies by Mas-Hesse \& Kunth (1999) reached a similar
conclusion. Using indicators of young stellar populations (the UV
continuum, the stellar + interstellar $W$(Si~{\sc iv})/$W$(C~{\sc iv})
ratio, the strength of the Wolf-Rayet bump), these authors conclude
that the majority of these galaxies have experienced recent nearly
instantaneous star formation dominating the ultraviolet light. Yet
their synthetic spectra are often below the observed visible
continuum, indicating the presence of an older stellar population.

Other causes may contribute to the behavior of [\oiii]/\hb\ and
\heia/\hb\ as a function of \whb\ as well. For example, \hii\ regions
may become gradually density bounded or the covering factor may
decrease with time.  A lower covering factor decreases the \hb\ 
luminosity and thus \whb. Allowing for leakage of ionizing photons
from the outskirts of the emitting region maintains [\oiii]/\hb\ and
\heia/\hb\ at a high level.  This is an attractive possibility, since
as nebulae expand, they become less and less opaque to ionizing
photons (for a homogeneous gas sphere, the mass that can be ionized
changes as  $Q_{\rm H}/n$).  Alternatively, it is possible that most of
the giant \hii\ regions in the samples are density bounded or have a
covering factor smaller than 1, {\em regardless of their evolutionary
  state}. Another possibility is that the spectroscopic data encompass
only part of the nebula (for simplicity, we will refer to these
explanations as the ``aperture effect'').  Such ``aperture effects''
are likely to occur in our \hii\ galaxy samples. A detailed study of
the \hii\ galaxy I~Zw~18 indicates that the brightest \hii\ region,
which dominates the spectrum, is density bounded, at least in some
directions (Stasi\'nska \& Schaerer 1999).  This could be a general
property since diffuse or filamentary ionized gas has been observed at large
distances in a number of \hii\ galaxies (e.g., Hoopes et al.\ 1996;
Martin 1997; Hunter \& Gallagher, 1997), implying that the data in this paper may be missing part
of the distant low-ionization gas.

High values of [\oiii]/\hb\ and \heia/\hb\ at low \whb\ could also
result if the extinction of the ionized gas is systematically higher
than that of the stellar light, as found in several studies of \hii\ 
galaxies and starbursts (e.g.\ Fanelli et al.\ 1988, Calzetti 1997,
Mas-Hesse \& Kunth 1999, Schaerer et al.\ 2000).

To summarize, simple photoionization models of \hii\ galaxies as pure
evolving starbursts cannot entirely account for the observed diagrams involving
\whb\ (panels a -- d) of Fig 1 -- 3. One or more of these effects are
likely explanations: (i) an older stellar population; (ii) starlight
systematically suffering a smaller extinction than ionized gas; or
(iii) an aperture effect where the observed spectra correspond to
density bounded nebulae or nebulae with a covering factor smaller than
1.  These interpretations explain in a natural way the absence of
\hii\ galaxies with \whb\ larger than 400 \AA\ in objective prism
surveys, whereas starburst models for ionization bounded \hii\ regions
predict values of up to 700 \AA\ (see also discussion in SL96). This
also implies that such surveys are biased against galaxies where the
latest starbursting episode is older than about 5~Myr, as such objects
would tend to have too weak \hb\ and \Ha\ emission lines to be
detected.

How can these explanations be reconciled with the increase of
[\oi]/\hb\ as \whb\ decreases (panels c in Figs.\ 1~--~3)? This trend
is also present in the corrected template spectra from Raimann et al.\ 
(Fig.~13), but the number of points is small and confirmation on a
larger sample is required.  One could argue that the old underlying
population is stronger in higher metallicity objects, so that the
increase in [\oi]/\hb\ with decreasing \whb\ could be attributed to a
metallicity effect.  We can test this idea by plotting
[\oi]/([\oii]+[\oiii]) as a function of O/H for the 60 objects with
measured \oiiia\ of the Izotov sample. This is done in panel b of
Fig.~14. Panel a of this figure shows the same objects in the
[\oi]/([\oii]+[\oiii]) vs. \whb\ plane.  Clearly, the metallicity
effect is not a good explanation.  Also, the behavior of [\oi] cannot
be explained by the ``aperture effect'' discussed above.  Reducing the
covering factor while keeping the nebulae ionization bounded does not
change the [\oii]/\hb\ and [\oi]/\hb\ ratios.  Making the nebulae
density bounded reduces the ratios.  Note that the observed behavior
of [\oi]/\hb\ with \whb\ argues in favor of \whb\ remaining a
statistical age indicator (although the age cannot be obtained by a
direct comparison with photoionization models for pure starbursts).
As pointed out by by Stasi\'nska \& Schaerer (1999) in the case of I
Zw 18, strong [O~{\sc i}] emission is easily produced by
photoionization models in dense filaments which may contribute only
marginally to the total emission in the other lines. Possibly, giant
\hii\ regions are globally density bounded but contain high density
filaments. Such filaments could be produced by shock-wave compression
or instabilities, produced by stellar winds and supernovae whose
effects are expected to increase with time.

\subsection{On the strong line method to derive oxygen abundances}

An important consequence of these observational selection effects in
\hii\ galaxy samples is that the age of the ionizing population has
little influence on the observed ([\oiii]+[\oii])/\hb\ ratio (see
panel e in Fig.~6~--~9). Combined with the narrow range of $U$ in
giant \hii\ galaxies (cf. panels a of Figs.\ 1~--~3 and Figs.\ 
6~--~11), this explains a posteriori why strong line methods based on
([\oiii]+[\oii])/\hb\ and methods utilizing the electron temperature
lead to similar oxygen abundance estimates in these objects (Pilyugin
2000).  A priori one would have expected large differences induced by
the temporal variation of the ionizing radiation field during the
evolution of a starburst.

Fig.\ 15 shows how the [\oiii]/[\oii] vs ([\oiii]+[\oii])/\hb\ diagram
(introduced by McGaugh 1991, 1994 to derive O/H and $U$ at the same
time) changes over a period of 4~Myr.  Continuous lines join models
with equal metallicities (the thickest lines correspond to the most
metal rich cases), dotted lines join models with equal $M_*$.  Panels
a, b, c correspond to starburst ages of .01, 2.01 and 4.01~Myr,
respectively. The objects from the Terlevich sample are overlaid in
the three panels.  We see that the theoretical diagrams at
metallicities lower than half solar evolve little over this period,
demonstrating the applicability of the strong line method at these
metallicities. In practice, it is however preferable to calibrate the
method not on models but empirically with a sample of high
signal-to-noise observations of \hii\ galaxies, as done by Pilyugin
(2000).

Interestingly, as already pointed out by Stasi\'nska (1999), rapid
evolution of the McGaugh (1991) diagram is, however, predicted at
metallicities $\ga$ solar, due to important changes in the hardness of
the stellar radiation field at such metallicities, {\em and} due to the
strong impact on optical line ratios.  This behavior of the stellar
radiation field at large metallicity is due to the fact that the overall main
sequence is shifted toward lower temperatures, and thus the
temperature of the stars falls below $T_{\rm eff} \la$ 40
kK earlier, as the hardness measured by $Q_{\rm He}/Q_{\rm H}$ decreases
rapidly.  This explains the more rapid decrease of the hardness of the
ionizing flux of stellar populations at high metallicity (cf.\ 
Schaerer \& Vacca 1998, Leitherer \etal\ 1999)\footnote{Although in
  general the hardness decreases with age, the temporary increase in
  the high metallicity models at 4 Myr is due to the appearance of WR
  stars at $\ga$ 3 Myr (cf.\ Fig.\ \ref{fig_q1q0}).  The reality of
  this feature is, however, questionable (cf.\ Section
  \ref{s_models}).}.  Therefore at metallicities around solar or above
(where electron temperature based methods cannot be used), the results
from strong line methods to derive the oxygen abundance are likely to
be rather inaccurate.

We also emphasize that strong line methods are statistical, as
compared to electron temperature methods. They should be used with
caution in cases when a systematic variation in the starburst
properties or the gas density distribution is expected. This applies
to studies of environmental effects on metallicities, e.g., such as
those of giant \hii\ regions seen in tidal tails of galaxies (Duc \&
Mirabel 1998) or \hii\ galaxies in the core of galaxy clusters (V\I
lchez 1995) as compared to isolated \hii\ galaxies

\subsection{Pure emission line ratio diagnostic diagrams}

Next we consider classical diagnostic diagrams which exclusively rely
on line ratios. To a first approximation, such diagrams depend only on
the population of the most massive stars which produce the ionization.
They have been extensively studied for giant \hii\ regions in spiral
galaxies and, as mentioned in Section~2.4, it has been found that
[\oiii]/\hb\ versus [\oii]/\hb, or [\oiii]/\hb\ versus [\oii]/[\oiii]
define a very narrow sequence, called the \hii\ region sequence. This
sequence was analyzed with the help of photoionization models and has
been interpreted as a sequence in metallicity with effective
temperature/ionization parameter varying in line with metallicity
(McCall et al.\ 1985, Dopita \& Evans 1986, Dopita et al.\ 2000).

Since \hii\ galaxy samples are believed to contain mostly
low-metallicity objects, it is not surprising that the \hii\ galaxy
sequences shown in Figs.\ 1~--~3 a are less populated at the low
[\oiii]/\hb\ end than the McCall et al.\ (1985) or van Zee et al.\ 
(1998) sequences, and that they extend to high [\oiii]/\hb.

Although some of the observed trends are met by the predictions, it
is, nevertheless, apparent from Figs.\ 6~-~11 that our models do not
reproduce the sequence very well.  Compared to other classical
diagnostic diagrams, the [\oiii]/\hb\ vs [\oii]/\hb\ diagram has the
advantage of being independent of the abundances ratios of O/N or O/S.
Therefore we will focus our discussion on this diagram, but the
conclusions are similar for the [\oiii]/\hb\ vs [\sii]/\hb\ diagram.
Diagrams involving [\nii] are discussed in Sect.\ \ref{s_no}.

Excluding the 2~\Zs\ and \Zs\ sequences (models represented by squares
and circles), which are not representative of the bulk of the objects in our samples,
the parameter space occupied by the models in the diagram does not
exactly match the observations.  First, our model sequences predict an
almost vertical drop in the [\oiii]/\hb\ vs. [\oii]/\hb\ diagram, and
the maximum [\oii]/\hb\ is almost independent of the ionization
parameter. This is due to oxygen mainly being in the form of O$^{+}$
when the most massive stars have disappeared.  Since we have argued
above that our \hii\ galaxies sample should contain few, if any,
galaxies with the most recent starburst older than 5~Myr, our
observational diagrams should be compared to the model sequences until
5~Myr only (roughly corresponding to \whb\ $\ga$ 50 \AA\ in Figs.\ 
6~--~11).

Even with that restriction, our photoionization models are unable to
reproduce [\oii]/\hb\ ratios larger than 4 and the kink at
[\oiii]/\hb\ $>$ 4 and [\oii]/\hb\ $>$ 2.  Moreover, the problem may
become even worse if the model spectra used for the Wolf-Rayet phase
were too hard (cf.\ Section \ref{s_models}).
 
We have computed supplementary photoionization models especially for
this purpose (including depletion of metals on grains or heating by
X-rays), but without success. The disagreement becomes even larger if
we assume that \hii\ galaxies are density bounded, as argued above.
Whether additional heating sources (shocks, conduction, turbulence)
may solve the problem is an open question.

Dopita et al.\ (2000) compared an observed sample of giant regions in
spiral galaxies with photoionization models and found that their
models could not reproduce the high [\oi]/\hb\ and [\sii]/\hb\ ratios
seen in some objects, invoking shocks as an explanation.  Our
models use a different prescription for the density structure than
those of Dopita et al.  and differ also in other respects, such as
numerical aspects and the N/O vs. O/H prescription.  Still, we find
that the problems exists not only for [\oi]/\hb\ and [\sii]/\hb\ but
also for [\oii]/\hb. Of course, the observed [\oii]/\hb\ ratio is
largely affected by the dereddening procedure, while this is much less
the case for the other two ratios.  Since strong line diagnostics
heavily rely on [\oii], and since this line is becoming increasingly
important for the study of galaxies at high redshift, it would be
useful to better constrain the problem by a detailed observational and
theoretical study of those few \hii\ galaxies with a firm indication
of shock excitation.

McCall et al.\ (1985) argued that giant \hii\ regions must be mostly
ionization bounded, otherwise one would not observe such a tight
sequence in emission line ratio diagrams.  This is at variance with
the fact that we now know that a fair portion of Lyman continuum
radiation is leaking out of \hii\ regions (Martin 1997; Stasi\'nska \&
Schaerer 1999, Beckman \etal\ 2000).  However, if the density
structure of giant \hii\ regions were driven by the mechanical action
of winds and supernovae explosions from their embedded stellar
populations, one could understand why there is so little dispersion
among giant \hii\ regions in such diagrams, even if they are partially
density bounded, since the driving parameter would boil down to the
stellar population itself.  In this respect, hydrodynamic modelling of
the ionization structure of giant \hii\ regions (the follow-up of the
single star \hii\ region modelling performed by Rodriguez-Gaspar \&
Tenorio-Tagle 1998) would be extremely important.

\subsection{The N/O ratio}
\label{s_no}

Among others, Kunth \& Sargent (1986), Pagel et al.\ (1986, 1992), Olofsson (1995b)
suggested that gas might be chemically enriched in, e.g., helium and
nitrogen from the winds of Wolf-Rayet stars during the evolution of
giant \hii\ regions.  Traces of such a local pollution have been found
in the irregular galaxy NGC~5253 from imaging spectroscopy (Walsh \&
Roy 1987, 1989, Kobulnicky et al.\ 1997), but other attempts to detect
local N/O enhancement in Wolf-Rayet galaxies have failed (Kobulnicky \&
Skillman 1996, 1998, Kobulnicky 1999). Moreover, Izotov et al.\ (1997)
and Kobulnicky \& Skillman (1996) analyzed observational data on \hii\ 
galaxies in the literature and found that galaxies with strong
Wolf-Rayet features in their integrated spectra exhibited the same N/O
ratios as the remaining galaxies at identical O/H ratios, contrary to
the claim of Pagel et al.\ (1986, 1992).

In the light of these previous studies, the [\nii]/[\oii] vs \whb\ 
diagram is revealing. There is a definite trend of [\nii]/[\oii]
strongly increasing as \whb\ decreases in all our \hii\ galaxy
samples. If we take [\nii]/[\oii] as an indicator of N/O, it is
tempting to attribute this trend to an increase in N/O as the ionizing
starburst gets older.  However, several biases must be examined first.
Could the observed relation be the result of a selection effect? While
the absence of low [\nii]/[\oii] ratios at small equivalent widths
could be attributed to selection effects against the weakest lines, at
large \Hb\ equivalent widths, values of [\nii]/[\oii] larger than
$\sim$ 0.2 -- 2 should be observed if such objects exist. Moreover,
there is no selection effect in the Izotov sample since all the
objects appear in Fig.\ 2f, and the relation is seen there as well.
The trend is less distinct than in the other samples, probably because
the sample is weighted towards the most metal poor galaxies. Finally,
as mentioned in Sect.~2.4, if a complete subsample is extracted from
the Terlevich sample, the observed relation between [\nii]/[\oii] and
\whb\ remains. Therefore, the observed relation cannot be attributed
to a simple selection effect.

Turning to an interpretation, it is important bearing in mind that
[\nii]/[\oii] does not only depend on N/O but also on the electron
temperature. It is larger at lower electron temperatures and therefore
at higher O/H for a given N/O. Second, at metallicities above solar,
the [\nii] and [\oii] lines are produced by recombination rather than
by collisional excitation (if the N$^{++}$/N$^+$ and O$^{++}$/O$^+$
ratios in the nebula are not close to zero). Our models (panels f in
Figs.~6~--~11) do account for that. We have already argued that models
with ages $\ga$ 5~Myr may not be relevant to our samples of \hii\ 
galaxies, but that \whb\ is an indicator of the age of the ionizing
starburst (in the sense that smaller \whb\ correspond to larger ages).
We have also argued that one should perhaps consider density bounded
rather than ionization bounded models. However, the [\nii]/[\oii]
ratio would remain unchanged in a density bounded model.  The large
values of [\nii]/[\oii] ($\ga$ 0.3) seen in some of the \hii\ regions
of our samples therefore indicate a N/O ratio larger than the largest
predicted by our models (0.1 at Z = 2 \Zs).

If the metallicities (i.e., O/H ratios) in our samples have identical
distributions in each age bin, the observed trend could be explained
if the two following conditions hold at the same time: (i) the
relation between N/O and O/H is steeper than the relation based on
McGaugh (1991) adopted in our models and (ii) the most metal rich
objects are those whose \whb\ values are the most affected by the
contribution of an old stellar population.  The first hypothesis might
indeed be true. There are indirect indications, from studies of giant
\hii\ regions in spiral galaxies, that at high metallicities N/O
increases more rapidly than N/O $\alpha$ (O/H)$^{0.5}$ adopted in our
models (cf.\ van Zee \etal\ 1998, Henry \etal\ 2000). The second
hypothesis appears quite reasonable.  Earlier generations of stars are
necessary to produce the bulk of presently observed O/H in more metal
rich \hii\ galaxies, and these earlier generations still contribute to
the continuum.  Qualitatively, the results of Raimann et al.\ (2000b)
support this picture.  We note however that the anticorrelation
between [\nii]/[\oii] and \whb\ subsists in the Raimann et al.\ 
templates even after correction for the old stellar population (Fig.
13f). If this effect is real, it calls for an additional explanation.

N/O might be increasing with time. Such an explanation would be more
compatible with the observed [\nii]/[\oii] versus
([\oii]+[\oiii])/\hb\ diagram in Figs.~1~--~3 i, in which the scatter
in ([\oii]+[\oiii])/\hb\ at a given [\nii]/[\oii] is extremely small.
Models with ages smaller than 5~Myr (panels i in Figs 6~--~11) show
about the same scatter, but in order to compare with observations a
convolution of the theoretical scatter with observational errors, and
a correction for reddening and underlying \hb\ absorption is first required.
  With N/O increasing with time, the theoretical scatter at a
given [\nii]/[\oii] becomes smaller.  The hypothesis of N/O increasing
with time can in principle be checked directly by plotting the N/O
determined by electron temperature based methods as a function of
\whb. We performed this test. No obvious trend is seen, but none of
the objects with large observed [\nii]/[\oii] in our samples has
\oiiia\ measured, so this test is not conclusive.

Deep spectroscopy for a detailed analysis of the stellar populations
in the objects showing high [\nii]/[\oii] ratios should be obtained.
High resolution emission line imaging and tailored photoionization
modeling should also be undertaken, in order to put the strongest
possible constraints on O/H and N/O in these objects.

\section{Conclusions}

We have reanalyzed the emission line properties of three large samples
of \hii\ galaxies taken from the literature (Terlevich et al.\ 1991;
Izotov and collaborators 1994-2000; Popescu \& Hopp 2000) with the aim
of studying the temporal evolution of these objects during the lifetime of the ionizing
stars ($\la$ 10 Myr).  We have constructed a series of diagrams using
observed emission line ratios and equivalent widths, and found
significant trends.

We interpreted these trends with photoionization models 
for integrated stellar populations.
This study extends the work of Stasi\'nska \& Leitherer (1996), by 
including objects with no direct metallicity determination
and using updated evolutionary tracks and atmospheric models. 

The changes brought about by the inclusion of the latest non-rotating
Geneva stellar evolution models and the {\em CoStar} non-LTE line 
blanketed atmosphere models including stellar winds for O stars
have been summarized in Sect.\ 3.1. (cf.\ Fig.\ 5).
These models have been successfully compared to observations
in a large variety of studies related to massive stars (see e.g.\
Maeder \& Meynet 1994, Stasi\'nska \& Schaerer 1997, Oey et al.\ 2000,
Schaerer 2000).
Despite this, the adopted models have also some known weaknesses,
e.g.\ the neglect of stellar rotation leading to modifications
of the stellar tracks (cf.\ Maeder \& Meynet 2000), a possible
overestimate of the hardness of the ionizing flux above $\sim$ 40 eV
(Oey \etal\ 2000), or the lack of line blanketing in WR atmospheres
(see Sect.\ 3.1).
Essentially, the main dependence of the photoionization models
on the stellar input physics enters through 
1) the evolutionary timescales and the temperatures of main sequence O 
stars, and 
2) the hardness of the He$^0$/H ionizing spectrum. It is important to note that,
while detailed outputs from photoionization models are sensitive to the exact stellar
input physics, our main conclusions summarized below are quite robust. 

The underlying principle of the comparison between models and observations
is that the metallicity
distribution is age-independent for sufficiently large and homogeneous
samples, despite the fact that the metallicities of individual objects
may be a priori unknown.

We have shown that \hii\ galaxies selected by objective-prism surveys
do not correspond to the simplistic view of instantaneous starbursts
surrounded by constant density, ionization bounded \hii\ regions.  The
observed relations between emission line ratios and \hb\ equivalent
width can be understood if most \hii\ galaxies contain older stellar
populations contributing, sometimes rather significantly, to the
observed optical continuum. This is in line with the recent findings
of Raimann et al.\ (2000ab) who, from a spectroscopic analysis on the
stellar features in very high signal-to-noise templates of \hii\ 
galaxies, showed that populations as old as 100~Myr and up to a few
Gyr are detectable in the spectra and significantly affect the
observed \hb\ equivalent widths. This is also consistent with the
model of a long lasting low level of star formation in I~Zw~18, as
suggested by Legrand (1999).  Two additional mechanisms may play a
role. First, stars seem to suffer a smaller dust obscuration in the
visible light than the ionized gas. Second, some Lyman continuum
radiation is probably leaking out of most of the nebulae encompassed
by the observing slits.

The combination of these effects reduces \whb\ with respect to the
value predicted for an instantaneous starburst surrounded by an
ionization bounded nebula.  Therefore objective-prism selected samples
of \hii\ galaxies are unlikely to contain significant numbers of
starbursts older than about 5 Myr. Older stages in the evolution of
starbursts must be selected from photometric surveys based on broad-
or narrow-band colors.

An interesting consequence of this selection effect is that the strong
line methods for deriving oxygen abundances work rather well in metal
poor \hii\ galaxies because there is no large mean effective temperature
spread. Under these conditions, a strong line method which is
calibrated on objects with electron temperature based oxygen
abundances determinations like that proposed by Pilyugin (2000), is
the most valid approach to derive oxygen abundances in low
signal-to-noise spectra of \hii\ galaxies.  One must, however, keep in
mind the statistical nature of this method and the fact that the
unknown stellar absorption in \hb\ provides a further source of
uncertainty.

We have shown that [\oi]/\hb\ uniformly increases with decreasing
\whb. This behavior suggests that this line ratio results from
dynamical effects that shape the nebula and whose importance increases
with time.  These dynamical effects could also be responsible for the
small range in ionization parameters that account for the observed
emission line trends.

The classical emission line ratio diagnostic diagrams such as
[\oiii]/\hb\ vs [\oii]/\hb\ imply a sequence in oxygen abundance and
ionization parameter. This suggestion was made earlier on the basis of
single-star photoionization models, but a physical interpretation
remains elusive.  Extra heating sources in addition to the Lyman
continuum radiation from massive stars seem necessary in order to
explain the largest observed [\oii]/\hb\ ratios.

The [\nii]/[\oii] ratio is shown to increase as \whb\ decreases.  The
observed trend is even stronger when considering the sample of high
signal-to-noise \hii\ galaxy templates of Raimann et al.\ (2000ab)
after correction for the old stellar population. We conclude that
either the relation between N/O and O/H is steeper than that adopted
in our models (N/O $\propto$ O/H$^{0.5}$) {\it and} the underlying
stellar population is stronger at higher metallicities, or the N/O
ratio increases with time on a time scale of $\sim$ 5~Myr. This last
option would support a scenario of self-pollution of giant \hii\ 
regions by nitrogen produced in situ.  High signal-to-noise
observations to {\it directly} uncover the old stellar populations and
the \hb\ contamination in {\it individual}\/ objects, especially those
with the highest [\nii]/[\oii] ratios, are required to settle this
important issue.
 
The possible dependence of the above findings on the various
main unknowns (underlying populations, escape of Lyman continuum photons,
non stellar heating sources, etc.) affecting our analysis have been 
discussed in the respective sections.
Luckily, not all the  problems raised in this study are interconnected, 
and they may be attacked from various angles. 
E.g.\ our claim of underlying populations being responsible for an
effective ``age bias'' against burst events with ages $\ga$ 5 Myr
and the puzzling behavior of [\nii]/[\oii]
can be tested by additional and more detailed stellar population studies 
of \hii\ galaxies along the lines of Raimann \etal\ (2000ab), 
using high signal-to-noise spectroscopy.
A purely ``stellar'' solution seems now clearly excluded for the problem
of [\oiii]/\hb\ vs [\oii]/\hb\ as well as [\sii]/\hb\ (Sect.\ 4.3 and
Stasi\'nska \& Schaerer 1999). 
Hydrodynamical models of the interaction of the interstellar medium
 with clusters of hot stars (in the vein of Cant\'o et al.\ 2000 or Franco et al.\ 2000) 
combined with photoionization calculations, should allow one to propose quantitative solutions. 
Such an approach may at the same time explain the behavior of [\oi]/\hb\ as a function of \whb\ 
and provide some quantitative estimate of the leakage of ionizing photons from giant \hii\ regions.
Multi-wavelength high spectral resolution observations should provide important constraints
 to the models in this respect.
The dust obscuration issue is perhaps trickier, since dust effects are extremely dependent on geometry.
Here, systematic high resolution imaging of giant \hii\ regions,
 such as performed by Calzetti et al.\ (1997) or Johnson et al.\ (2000), 
 should allow one to clarify the 
dust location and its effect on integrated \hii\ galaxies spectra. 
The insight gained from this study and the proposed directions for further 
theoretical and observational investigations
will considerably increase our understanding of the physics of \hii\ 
regions and the complex interactions between its stellar and interstellar
components.

\acknowledgements

We thank E. Telles for sending us the list of objects in the Terlevich
sample that proved to be giant \hii\ regions in spiral galaxies.
We thank D. Raimann for sending us unpublished data that were
required to construct Fig.\ 13.


\begin{figure*}
\centerline{\psfig{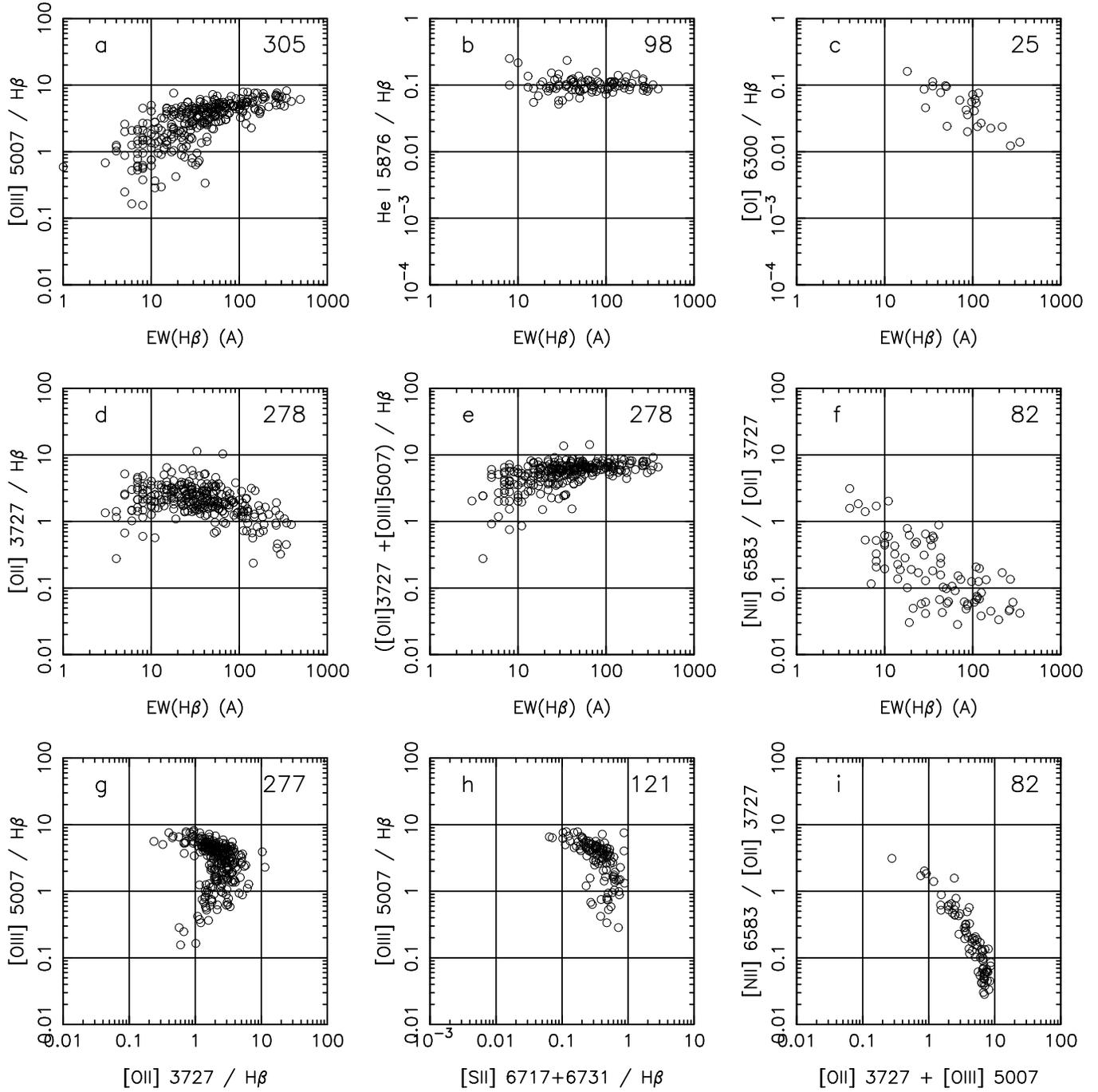}}
\caption{The Terlevich
sample of \hii\ galaxies.
On the upper right of each panel is given the total number of the objects 
appearing in the diagram.}
\end{figure*}

\begin{figure*}
\centerline{\psfig{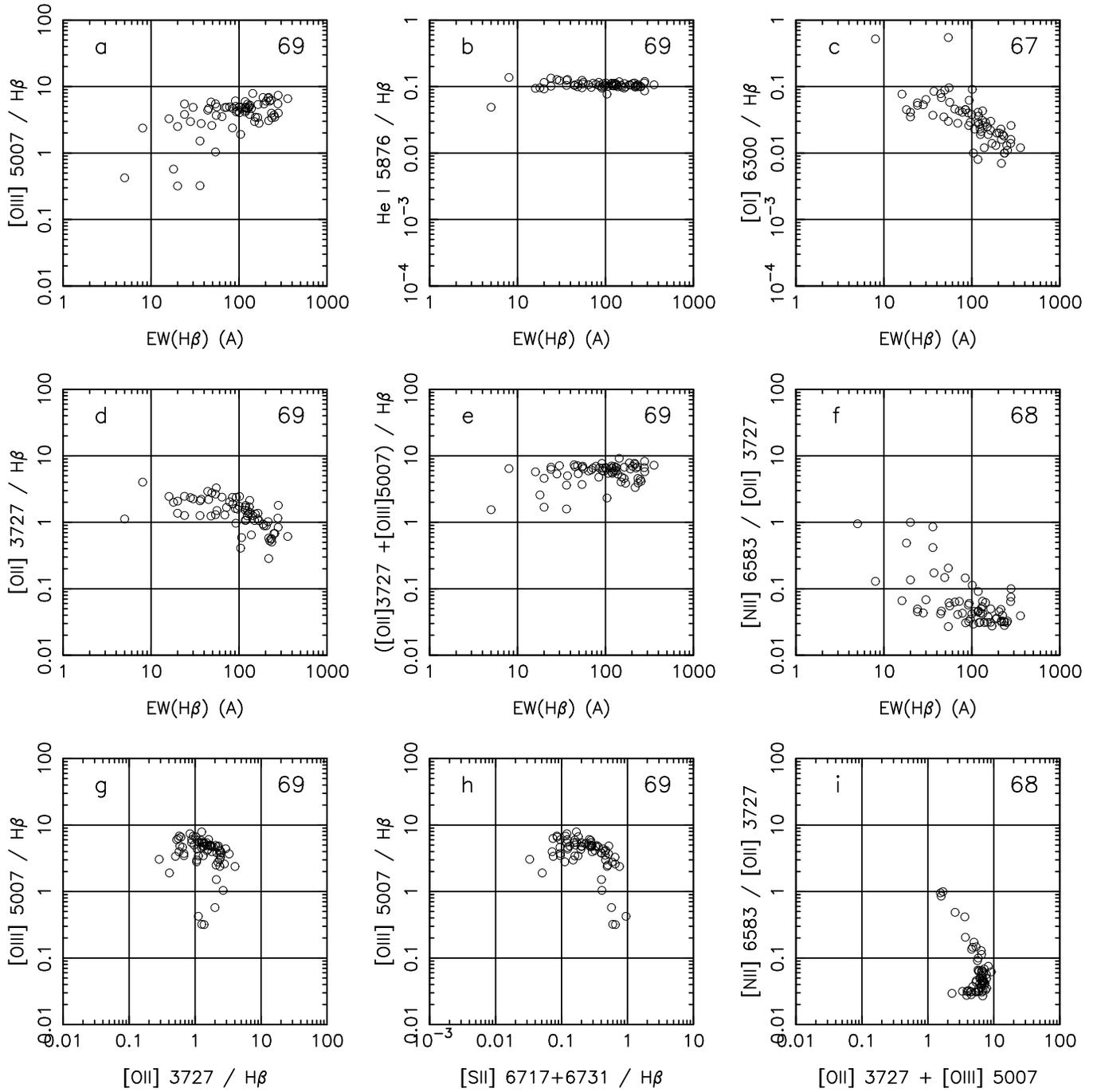}}
\caption{Same as Fig. 1 for the Izotov sample of \hii\ galaxies.}
\end{figure*}

\begin{figure*}
\centerline{\psfig{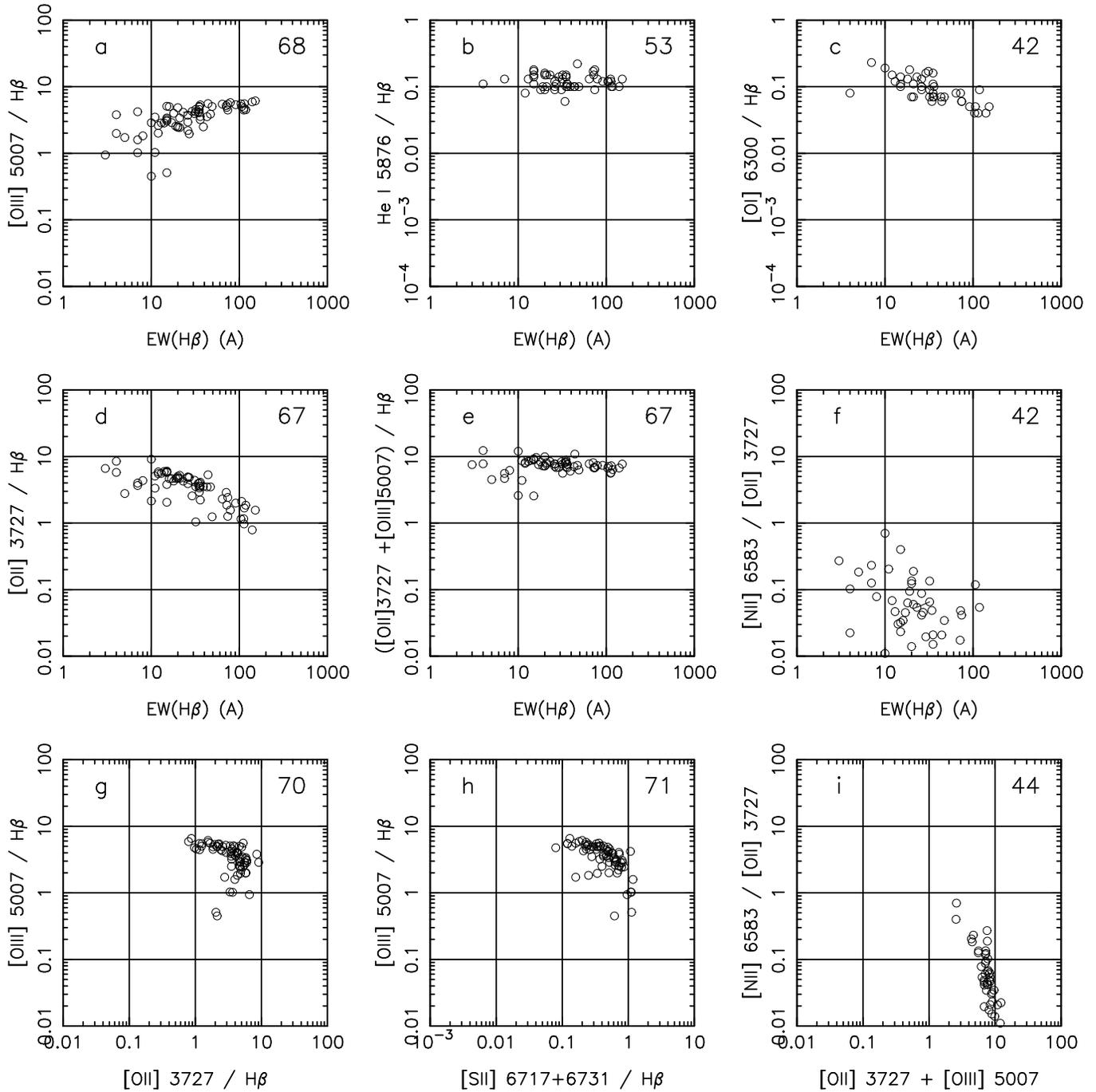}}
\caption{Same as Fig. 1 for the Popescu \&
Hopp sample of \hii\ galaxies.}
\end{figure*}

\begin{figure*}
\centerline{\psfig{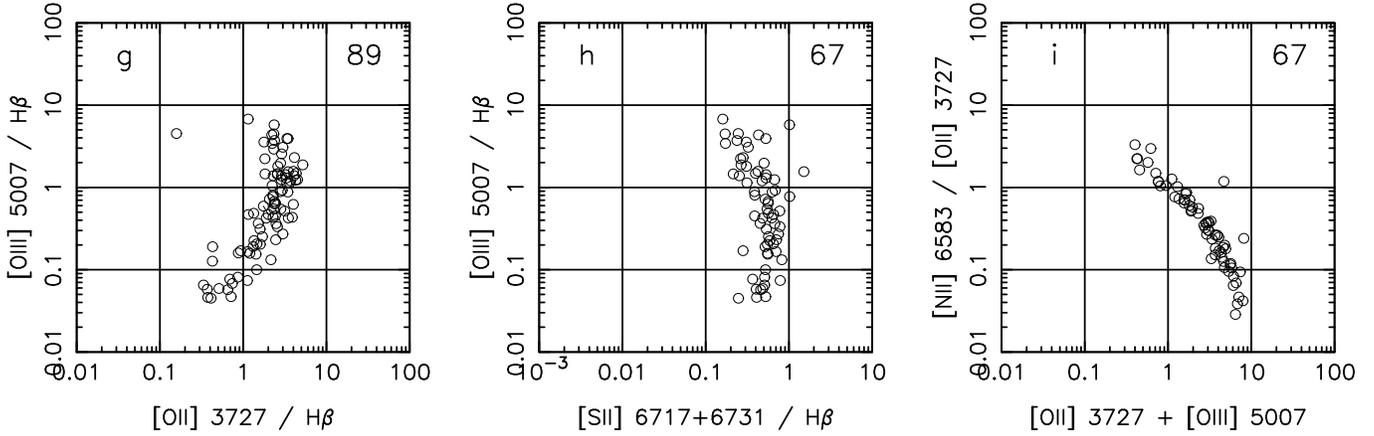}}
\caption{Emission line ratio
diagrams for the sample of giant \hii\ regions in spiral galaxies from Mc Call et al.\ (1985).}
   \end{figure*}

\begin{figure*}
\centerline{\psfig{figure=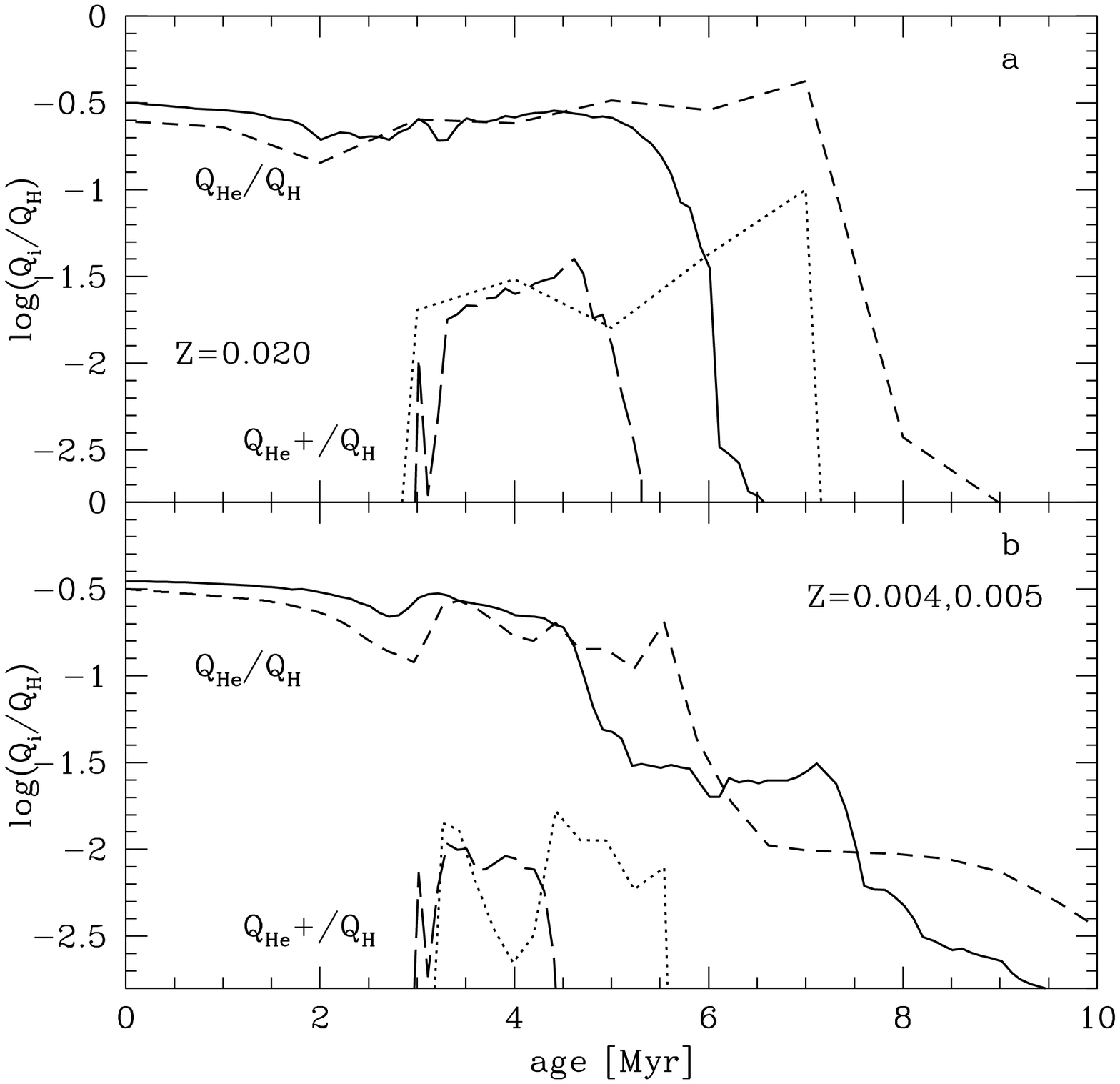,width=8.8cm}
\psfig{figure=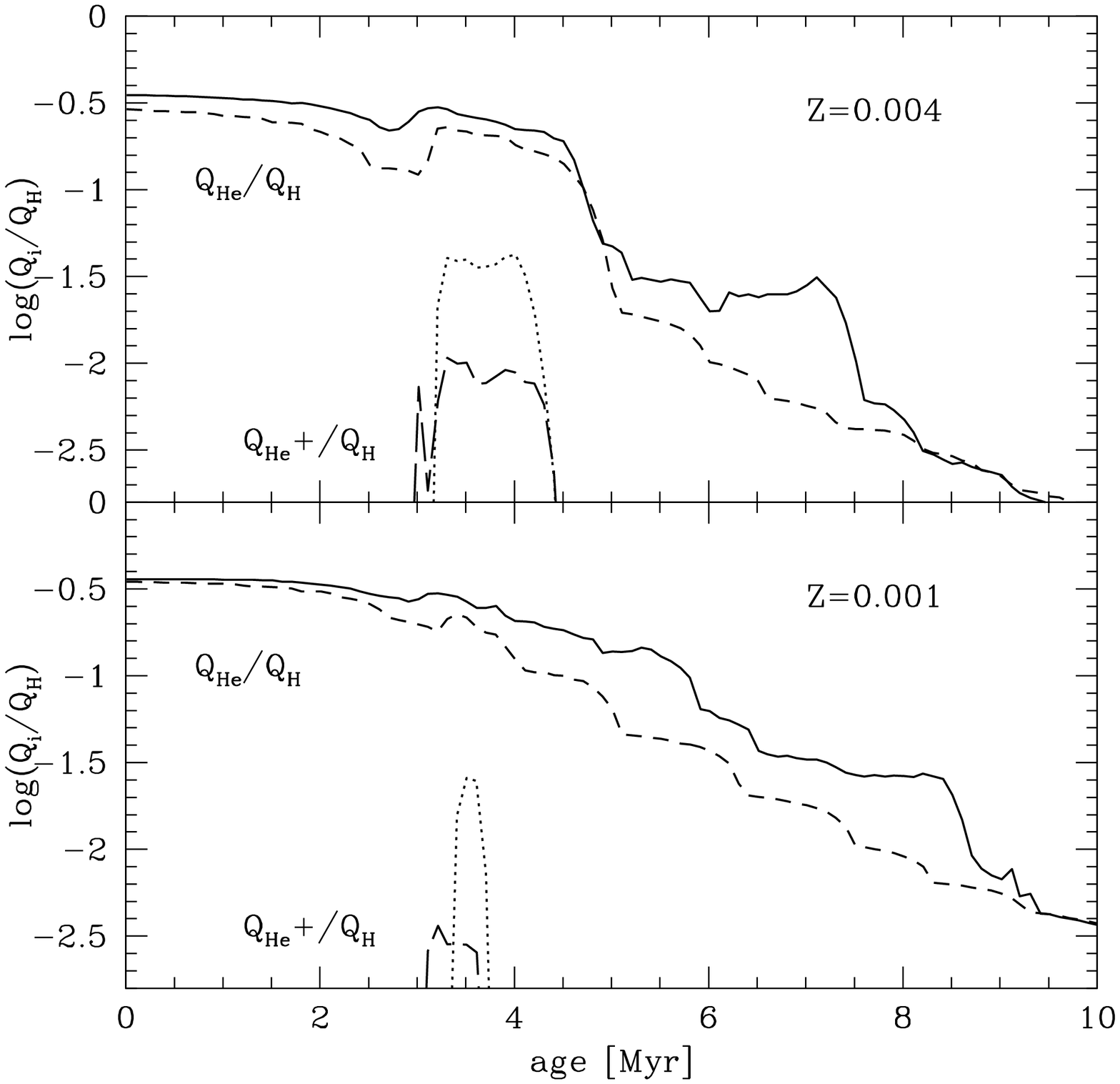,width=8.8cm}}
\caption{Comparison of the predicted hardness of the ionizing spectra
from various evolutionary synthesis models for instantaneous bursts, 
Salpeter IMF, and different metallicities.
Illustrated are the ratios of the He$^0$ and He$^+$ to H ionizing photon fluxes 
($Q_{\rm He}/Q_{\rm H}$ and $Q_{\rm He^+}/Q_{\rm H}$ respectively).
{\bf a)} Schaerer \& Vacca (1998, hereafter SV98) models (solid, long-dashed) versus
Leitherer \& Heckman (1995) models (dashed, dotted) at solar metallicity.
{\bf b)} SV98 models (solid, long-dashed) at 1/5 solar,
Leitherer \& Heckman (1995) models (dashed, dotted) at 1/4 solar.
{\bf c)} SV98 models (solid, long-dashed) versus 
{\em Starburst99} models (dashed, dotted) at 1/5 solar.
{\bf d)} Same as c for 1/20 solar metallicity.
Discussion in text.}
\label{fig_q1q0}
\end{figure*}

\begin{figure*}
\centerline{\psfig{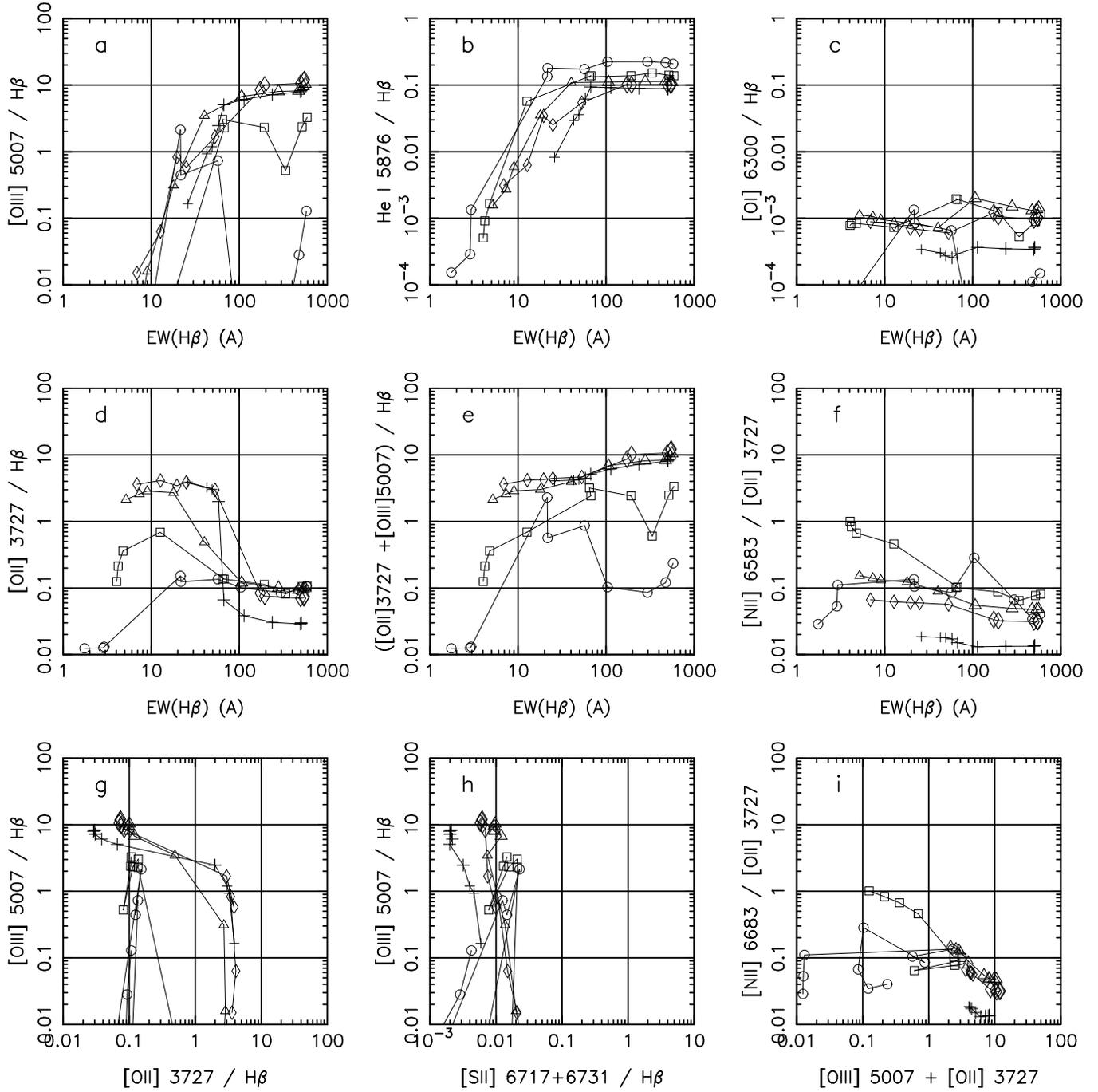}}
\caption{Sequences of models for
instantaneous bursts (\mup\ = 120 \Ms, \mstar = 10$^9$ \Ms), full sphere geometry,
$n$ = 10 cm$^{-3}$, various metallicities. 
The metallicity is indicated by the following symbols: circle for Z/\Zs\ =
2, square for Z/\Zs\ = 1, triangle for Z/\Zs\ = 0.4, diamond for Z/\Zs\ =
0.2 and plus sign for Z/\Zs\ = 0.05. 
The symbols mark time
steps of 1 Myr.}
\end{figure*}

\begin{figure*}
\centerline{\psfig{figure=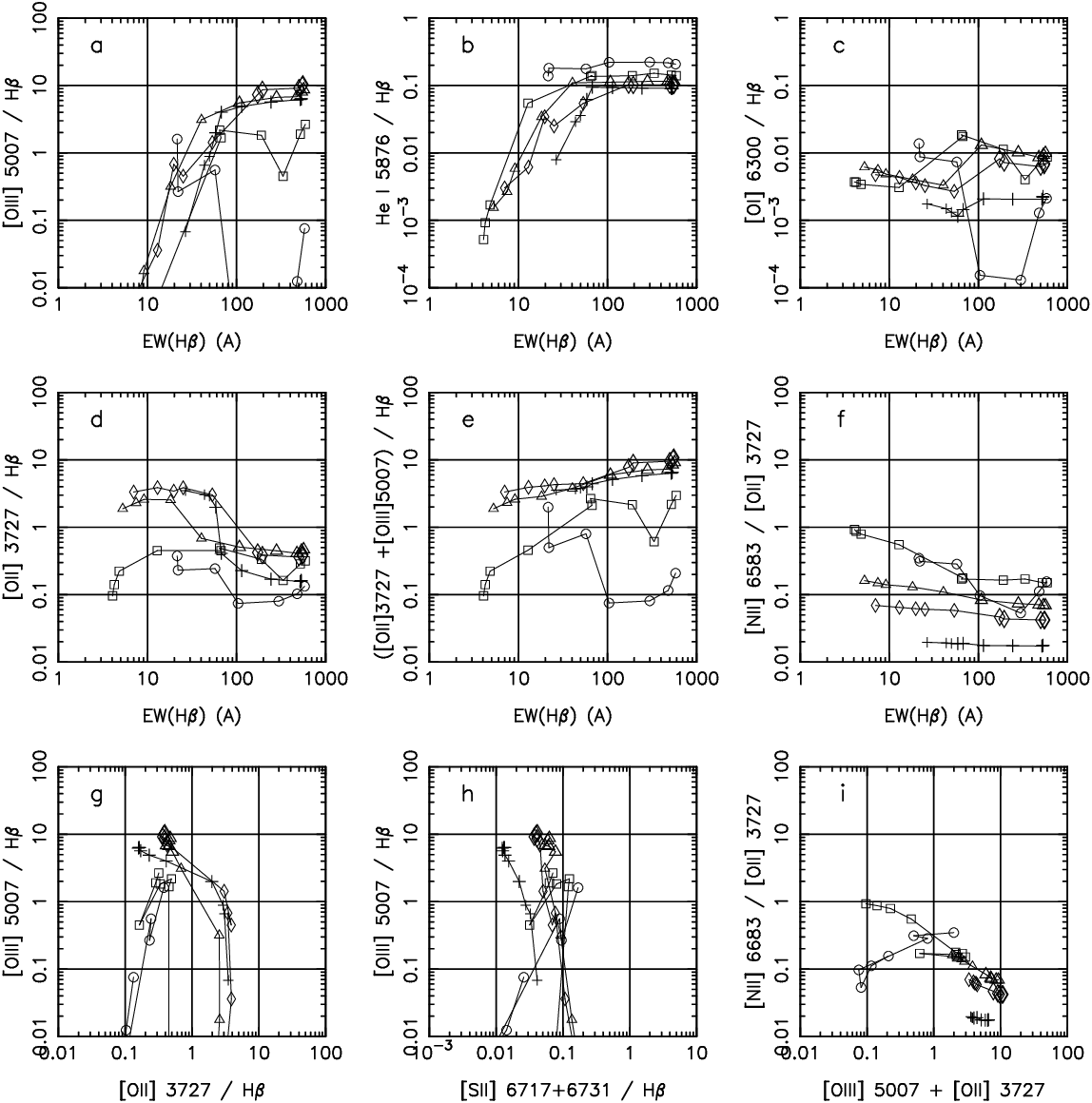,width=18cm}}
\caption{
Same as Fig. 6 but for  \mup\ = 120 \Ms, \mstar = 10$^6$ \Ms.}
\end{figure*}

\begin{figure*}
\centerline{\psfig{figure=HM_FIG_8.PS,width=18cm}} 
\caption{
Same as Fig. 6 but for  \mup\ = 120 \Ms, \mstar = 10$^3$ \Ms.}
\end{figure*}

\begin{figure*}
\centerline{\psfig{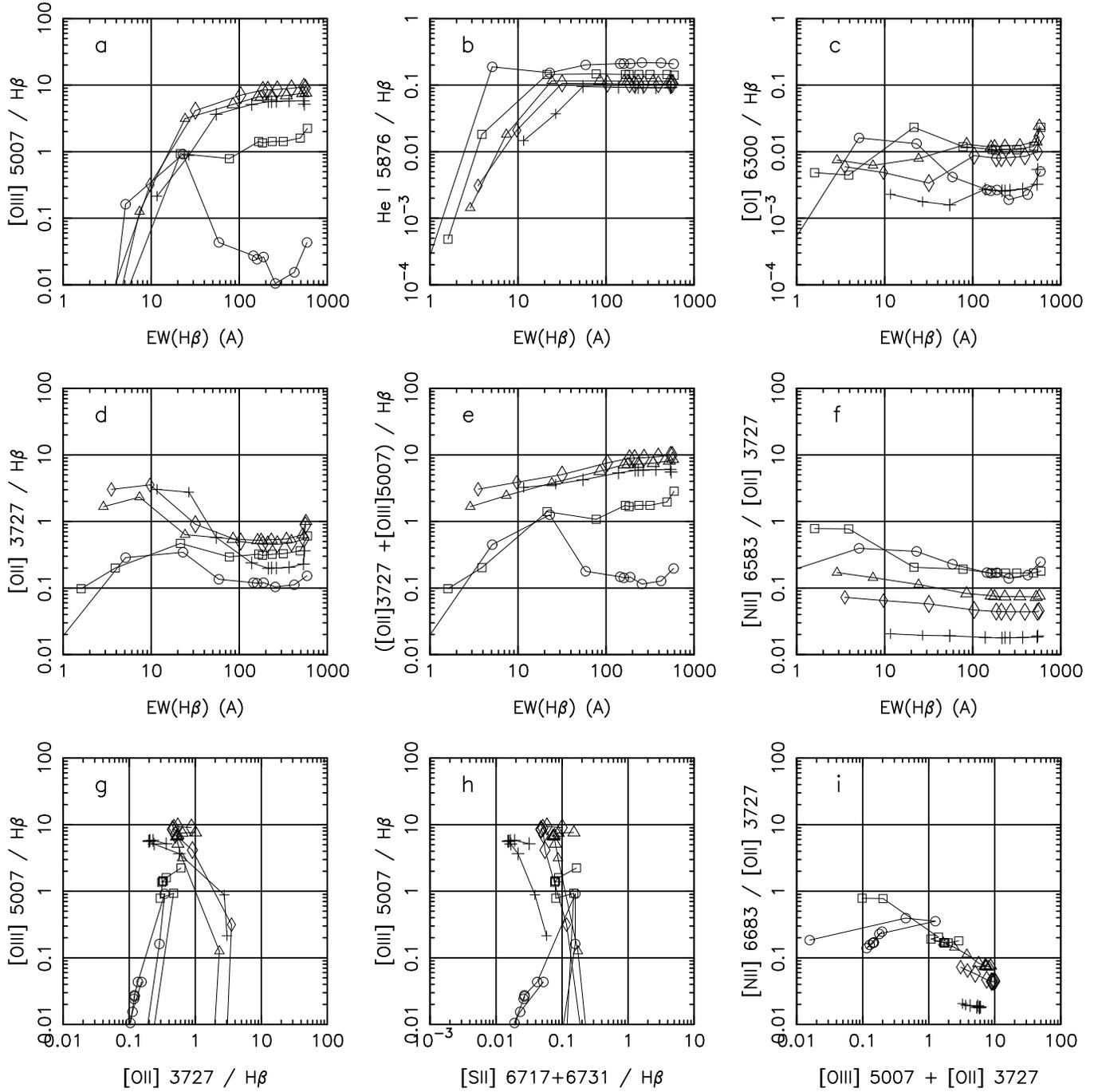}} 
\caption{
Same as Fig. 7, \mup\ = 120 \Ms, \mstar = 10$^6$ \Ms, but for an extended burst (duration 10
Myr). Same symbols as in Fig.\ 7 but with time
steps of 2 Myr.}
\end{figure*}

\begin{figure*}
\centerline{\psfig{figure=HM_FIG10.PS,width=18cm}} 
\caption{
Same as Fig. 6 but for  \mup\ = 30 \Ms, \mstar = 10$^6$ \Ms.}
\end{figure*}

\begin{figure*}
\centerline{\psfig{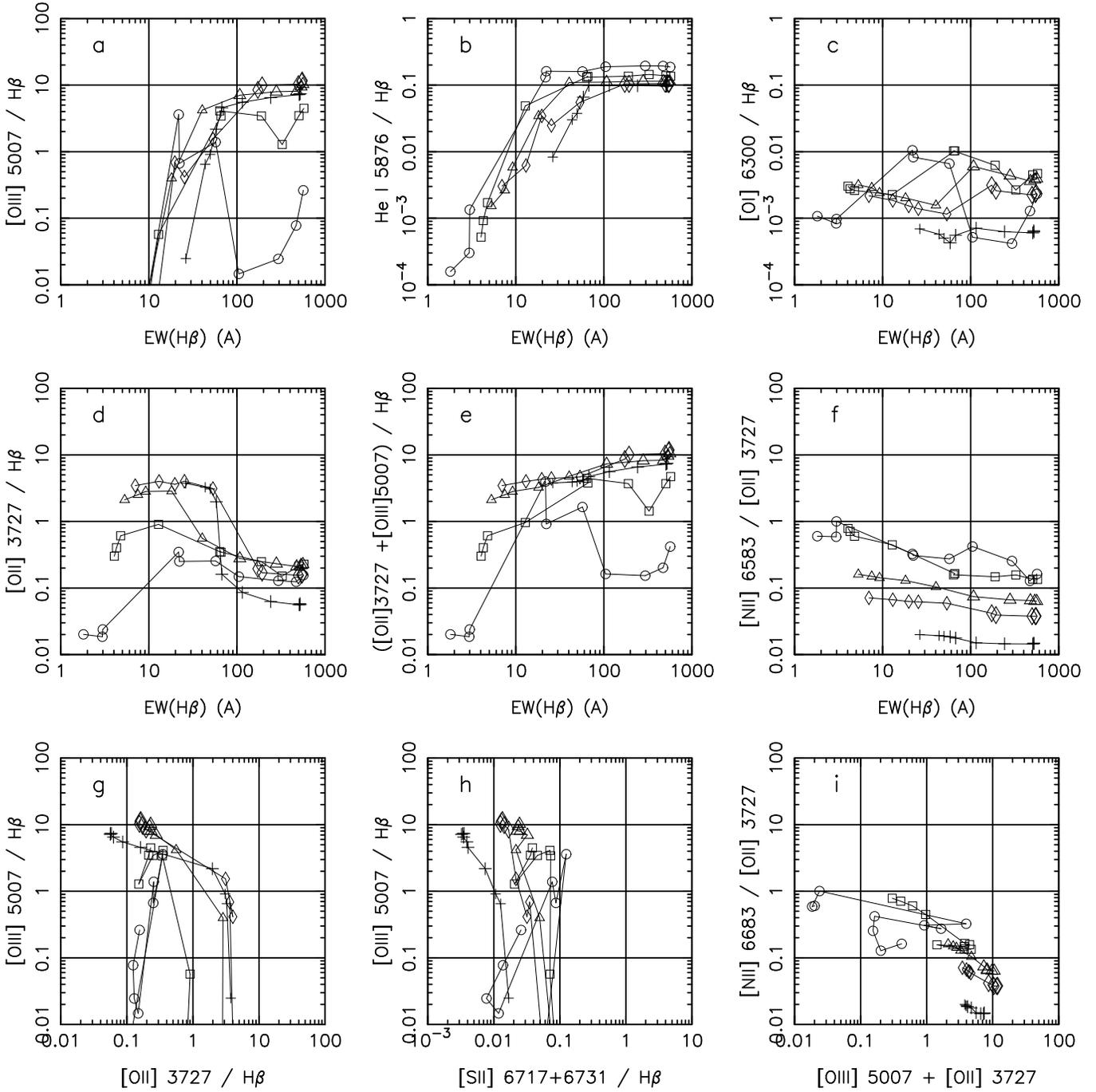}} 
\caption{
Sequences of models for
instantaneous bursts (\mup\ = 120 \Ms, \mstar = 10$^9$ \Ms), hollow sphere geometry,
$n$ = 200 cm$^{-3}$, various metallicities. 
These models have the same average ionization parameter as the models
of Fig.\ 7.
Same symbols as in Fig.\ 5.}
\end{figure*}

\clearpage
\begin{figure}
\centerline{\psfig{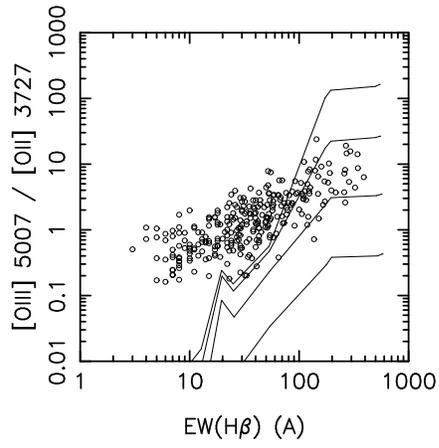}} 
\caption{Objects from the Terlevich sample superimposed on sequences of 
instantaneous burst models (\mup\ = 120 \Ms, \mstar\ = 1, 10$^3$, 10$^6$ and 10$^9$ \Ms),
full sphere geometry, $n$ = 10 cm$^{-3}$
 metallicity Z/\Zs\ =0.2}
\end{figure}

\begin{figure*}
\centerline{\psfig{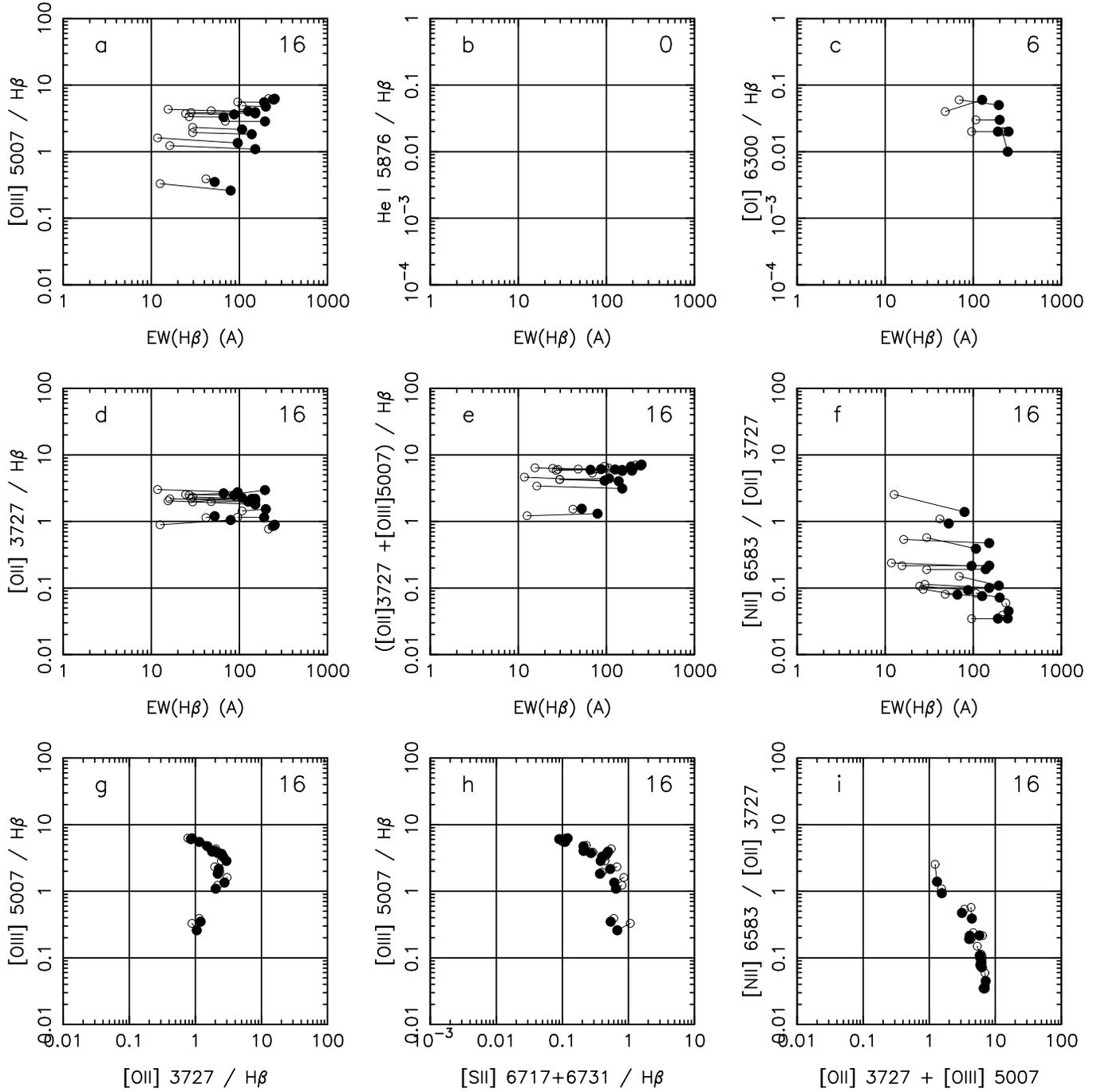}} \caption{ Groups of \hii\
galaxies as defined by Raimann et al.\ (2000a) from the Terlevich et
al. (1991) catalogue. Open circles: the averaged spectra as used by
Raiman et al.
(2000a);  filled circles: the same after correction for
underlying stellar population and internal reddening.
}
\end{figure*}

\begin{figure*}
\centerline{\psfig{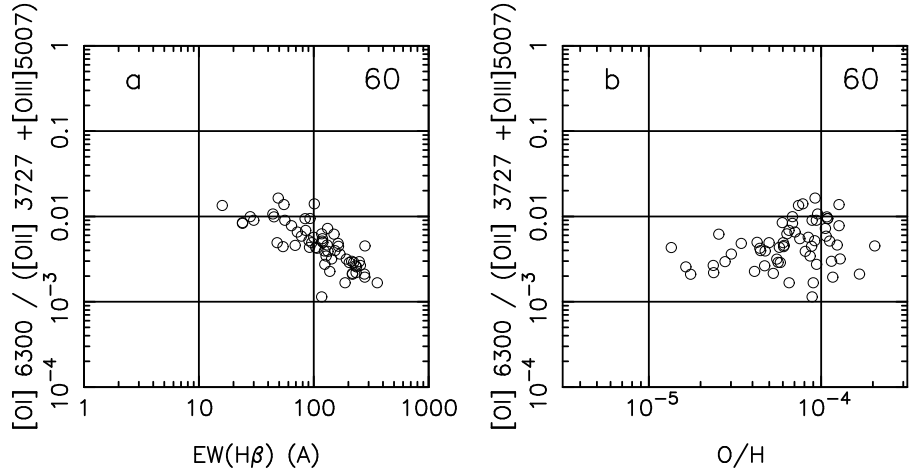}} \caption{Comparison of the trends 
for [\oi]/([\oii]+[\oiii]) as a function of \whb\ and O/H 
for the 60 objects with measured \oiiia\
of the Izotov sample. 
}
\end{figure*}

\begin{figure*}
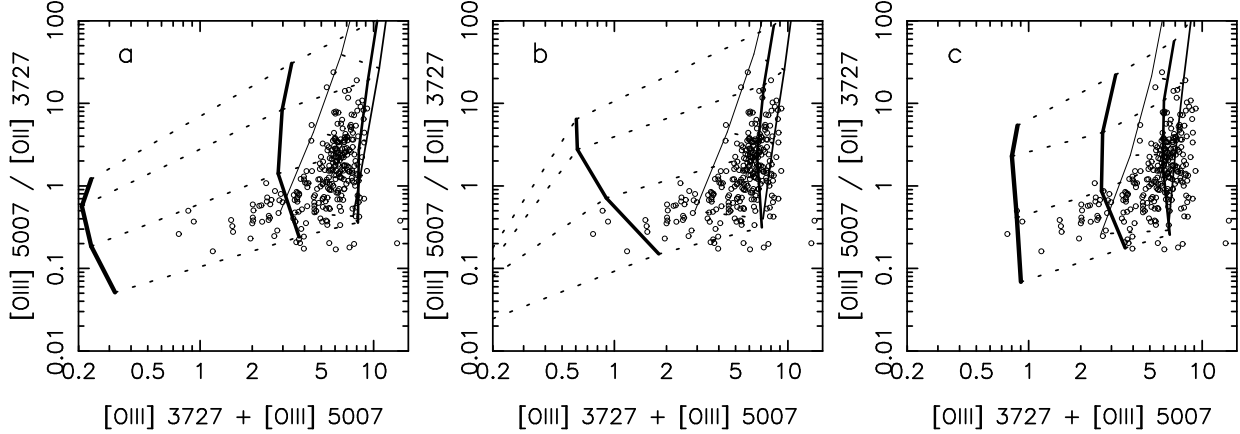

\centerline{\psfig{figure=HMO_FI15a.PS,width=5.4cm}
\psfig{figure=HMO_FI15b.PS,width=5.4cm}
\psfig{figure=HMO_FI15c.PS,width=5.4cm}}
\caption{Evolution of the McGaugh (1991) diagram over a period of 4 Myr
 for our instantaneous burst models
 (\mup\ = 120 \Ms), full sphere geometry,
$n$ = 10 cm$^{-3}$. 
Models with equal metallicities are linked by full lines 
(whose thickness is proportional to the metallicity).
Models with equal $M_*$ are linked with dotted lines.
Panels a, b, c correspond to starburst ages of .01,
2.01 and 4.01~Myr, respectively
Superimposed are the
objects from the Terlevich sample}
\end{figure*}

\end{document}